\documentclass{article}
\usepackage{PRIMEarxiv}

\usepackage[utf8]{inputenc} 
\usepackage[T1]{fontenc}    
\usepackage{hyperref}       
\usepackage{url}            
\usepackage{booktabs}       
\usepackage{amsfonts}       
\usepackage{amsmath}
\usepackage{nicefrac}       
\usepackage{microtype}      
\usepackage{lipsum}
\usepackage{fancyhdr}       
\usepackage{graphicx}       
\usepackage{comment}
\graphicspath{{media/}}     
\AtBeginDocument{%
  \providecommand\BibTeX{{%
    \normalfont B\kern-0.5em{\scshape i\kern-0.25em b}\kern-0.8em\TeX}}}

\usepackage{subfigure}
\usepackage{xcolor}
\usepackage{multirow}

\title{Identification of Adaptive Driving Style Preference through Implicit Inputs in SAE L2 Vehicles}

\author{
  Zhaobo K. Zheng, Kumar Akash, Teruhisa Misu \\
  Honda Research Insitute USA, Inc \\
  San Jose \\
  USA\\
  \texttt{\{zhaobo\char`_zheng, kakash,tmisu\}@honda-ri.com} \\
   \And
  Vidya Krishmoorthy, Miaomiao Dong, Yuni Lee, Gaojian Huang \\
  San Jose State University \\
  San Jose \\
  USA\\
   \texttt{\{vidya.krishnamoorthy, miaomiao.dong,yuni.lee,gaojian.huang\}@honda-ri.com} \\
}

\begin{document}
\maketitle

\author{
  Zhaobo K. Zheng, Kumar Akash, Teruhisa Misu \\
  Honda Research Institute USA, Inc \\
  San Jose \\
  USA\\
  \texttt{zhaobo_zheng,kakash@honda-ri.com} \\
   \And
  Vidya Krishnamoorthy, Miaomiao Dong, Yuni Lee, Gaojian Huang \\
  San Jose State University \\
  San Jose \\
  USA\\
  \texttt{gaojian.huang@sjsu.edu} \\
}

\begin{abstract}
  A key factor to optimal acceptance and comfort of automated vehicle features is the driving style. Mismatches between the automated and the driver preferred driving styles can make users take over more frequently or even disable the automation features. This work proposes identification of user driving style preference with multimodal signals, so the vehicle could match user preference in a continuous and automatic way. We conducted a driving simulator study with 36 participants and collected extensive multimodal data including behavioral, physiological, and situational data. This includes eye gaze, steering grip force, driving maneuvers, brake and throttle pedal inputs as well as foot distance from pedals, pupil diameter, galvanic skin response, heart rate, and situational drive context. Then, we built machine learning models to identify preferred driving styles, and confirmed that all modalities are important for the identification of user preference. This work paves the road for implicit adaptive driving styles on automated vehicles.\footnote{This is the author’s version of the work. It is posted here for your personal use. Not for redistribution. The definitive Version of Record was published in Proceedings of the 2022 International Conference on Multimodal Interaction (ICMI ’22), November 7–11, 2022, Bengaluru, India.}
\end{abstract}


\section{Introduction}
As automobile technology advances, more and more vehicles on the road are equipped with advanced driver assistance systems (ADAS), such as adaptive cruise control (ACC) and lane departure correction \cite{Knoop2019PlatoonStability}. These ADAS features could provide longitudinal and lateral control within operational limits, even though the drivers still need to maintain attention and physical readiness to intervene \cite{SAE2018TaxonomyInternational}. These features could bring improved safety, convenience and fuel efficiency on individuals as well as better traffic flow and reduced accidents on the society \cite{Casner2016TheDriving}\cite{Glumac2020PracticeLimitations}. Despite these great potentials, the acceptance and trust of ADAS features still remain a crucial factor, that users may turn off these features if they do not prefer the control behaviors and driving styles \cite{Akash2020TowardAutomation}. A recent study found out significant decreased user acceptance rates, when the driving style is different from the user preference, in terms of aggressive and defensive \cite{Ma2021DriversStyles}. Moreover, mismatch between ADAS and user preferred driving styles can cause more takeovers, leading to non-smooth drives or even unsafe situations \cite{Soares2021TakeoverMeta-analysis}. It is essential for ADAS features and future fully automated drives to adapt user preferred driving styles, to achieve higher acceptance, smoother interaction and improved driver comfort. 

Driving style is summarized as observable patterns of parameter sets related to the maneuver and trajectory planning \cite{Bellem2016ObjectiveVehicles} and the preferred driving style is a key factor to drivers' trust and comfort. Griesche et al. studied whether drivers would prefer a driving style similar to their own \cite{GriescheBosch2016ShouldAut}. While many drivers preferred a perfect matching driving style, there are still thirty percent of the drivers who preferred the opposite driving styles of their own. Hartwich et al. studied the effects of driving style familiarity and drivers' age on driving comfort and enjoyment \cite{Hartwich2018DrivingFamiliarity}. Younger drivers experienced the familiar driving style as more comfortable than the unfamiliar driving styles, but this effect was opposite for older drivers, who experienced more comfort with an unfamiliar driving style. Ekman et al. explored how the driving styles of automated vehicles would affect driver trust \cite{Ekman2019ExploringInformation}. The defensive driving style was perceived as more trustworthy, mainly because it deemed more predictable than the aggressive driving style. Phinnemore et al. investigated the effect of mood on preferred driving styles \cite{Phinnemore2021HappyCars}. They showed video stimulus to the drivers and found out that drivers are more likely to prefer aggressive, moderate and conservative driving styles under excited, neutral and calm moods. These existing studies demonstrate that a variety of factors could contribute to the preferred driving style, including driver affective state, situation and scenarios, which indicate that there is no unique driving style matched to each individual and optimal driving style may vary depending on situations.

Adaptive driving style has the potential to adjust driving styles according to these factors in a real-time and individualized manner, and there are some existing prototypes and research on it. For an example, current ACC functions usually allows users to adjust the controller to their preferred following distance from vehicle ahead \cite{Rudin-Brown2004BehaviouralStrategies}. More recently, Tesla has introduced self driving profiles, ranging from chill, average and assertive driving styles \cite{InsideEVs2022TeslaAssertive}, and users can choose their comfortable automated driving styles. These prototypes took the initial steps to adapt automated vehicles to preferred driving styles. However, these systems need explicit inputs on preferred driving style from users and it is inconvenient due to the frequent changeable nature of preferred driving styles. Despite some research trying to conveniently take the inputs through novel channels such as audio \cite{Kim2021GuidingStudy}, explicit adaptive driving styles still add extra workload on users. Furthermore, as the coverage of driving automation function expands from freeway to urban driving scenarios, further research is needed with the increased situation complexity and automation parameters.

To bridge this gap, this work proposes identification of preferred driving style through multimodal data in SAE L2 automated vehicles. We focus on L2 automation because it is the most popular automation level in the current society. We hypothesize that we can measure user comfort and satisfaction towards the current driving style, through multimodal behaviors. Then, we can identify their preferred driving style in a continuous and automatic manner, without explicit inputs on driving styles from the users. There has been existing research on detection of drivers' manual driving styles \cite{MarinaMartinez2018DrivingSurvey}, but this work focuses on the preferred driving style of the AV, which is how the drivers want to be driven. In summary, the contributions of this paper are as follows. 
\begin{enumerate}
    \item To the best knowledge of the authors, this work is the first in the field on implicit adaption of L2 automated driving style, that is even feasible in complicated urban driving. 
    \item We have integrated most existing and developed novel data modalities related to driver mental states, including driver behavior (eye gaze, maneuver, grip, pedal press and distance), physiological (pupil size and peripheral physiological), and situational (gaze semantics, drive) data. These sensing modalities are wearable and minimal-invasive, that could be later deployed in everyday drives. 
    \item We have conducted a user study on 36 participants in a driving simulator study to validate the feasibility of our identification framework. Based on the collected data, we performed statistical analysis to discover driver behavioral patterns under various driving style preferences, and trained machine learning models to automatically identify preferred driving styles through implicit inputs. 
\end{enumerate}
    
    The rest of this paper is organized as follows, Section~\ref{sec:user_study} describes the user study and data collection, including the experimental design and equipment. Section~\ref{sec:data_processing} introduces data processing, feature extraction, and model development with our data. Finally, Section~\ref{sec:stat} demonstrate the results of our analysis and followed by concluding statements in Section~\ref{sec:conclusion}. 

\section{User Study and Data Collection} \label{sec:user_study}
Thirty six participants (mean age = 22.86, standard deviation = 5.487, range:18-38) were recruited for this study, including 16 females, 19 males and 1 non-binary. All participants were required to have a valid driver's license and be within the age range of 18 to 65, with no self-reported sensory deficits. The entire experiment lasted approximately 2 hours, and participants were compensated with either a \$20 gift card or a two-hour (course) credit. The study was approved by San Jose State University Institutional Review Board (approval ID: 21232). We will make this dataset public available upon request for research purposes, upon the acceptance of this paper.

\subsection{Equipment}
The equipment used for this study mainly include a driving simulator and wearable sensors. The wearable sensors include an eye tracking glasses, a physiological sensor and a pressure sensing glove. 

\subsubsection{Driving Simulator}
\begin{figure}
\centering
\subfigure[Driving simulator appearance \label{fig:simulator}]{\includegraphics[width=.45\linewidth]{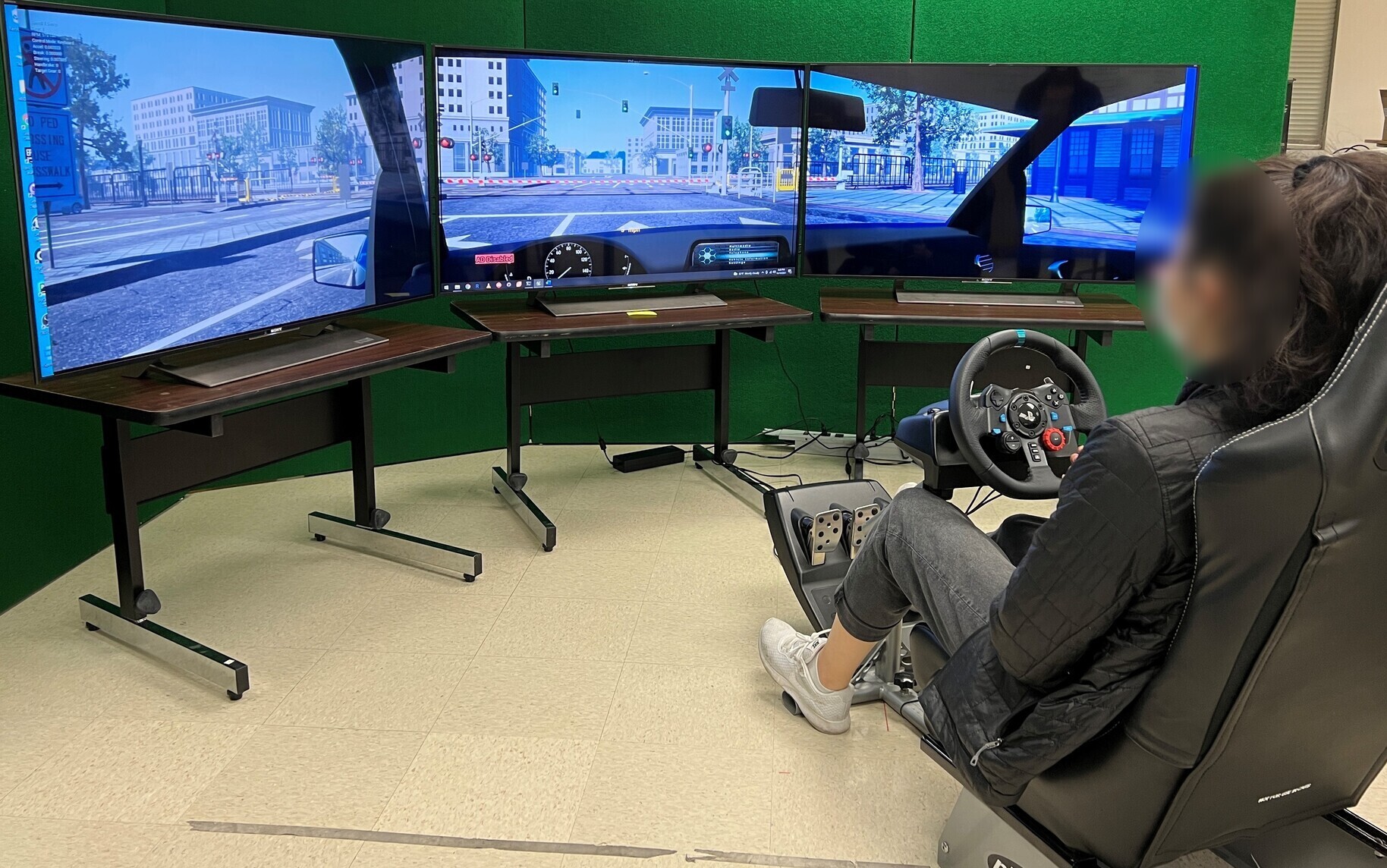}} \subfigure[Distance detection sensor on pedals \label{fig:pedals}]{\includegraphics[width=.375\linewidth]{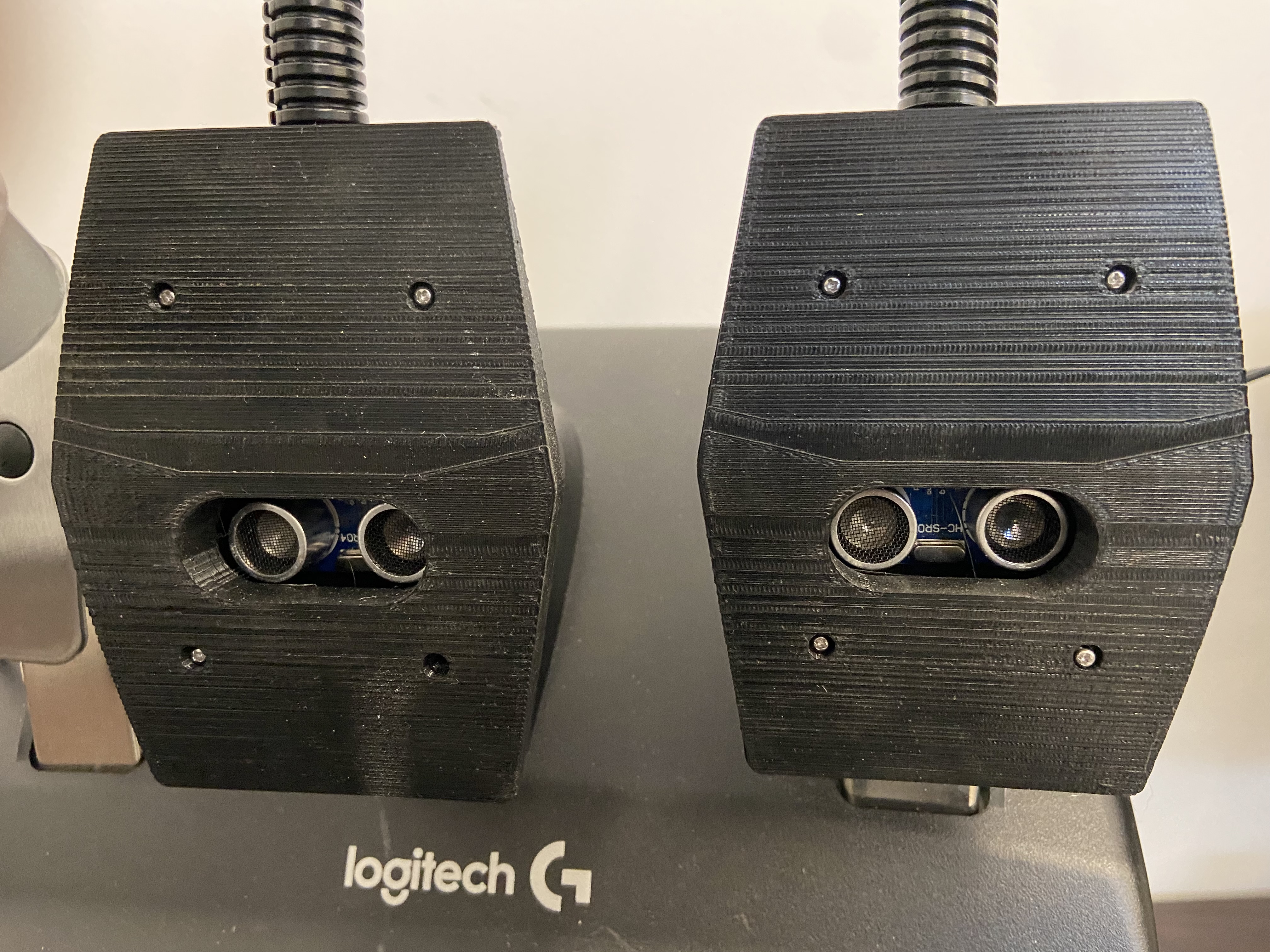}}
\caption{Driving simulator}
\label{drivingsimulator}
\end{figure}

The study was conducted using a medium-fidelity driving simulator which includes a Logitech G29 steering wheel, brake pedal, and accelerator pedal, as shown in Fig. ~\ref{fig:simulator}. The scenarios were shown on three 45-inch TV screens with a 50 degrees included angle and 42 inches perpendicular distance to the user. The user can see both the vehicle dashboard and the driving surrounding environment. The dashboard includes a speedometer showing the speed, a navigation arrow presenting the target direction (left/right/straight), and an indication of whether the driving automation is on or off. The driving environment  was rendered with the Unreal Engine \cite{EpicGames2022UnrealEngine}, simulating an urban area, which includes traffic lights, other vehicles, pedestrian crossing intersections, road signs and roundabouts, as shown in Fig. 2. 

To measure takeover intentions from participants, we integrated non-contact distance detection sensors on the throttle and brake pedals, as shown in Fig. ~\ref{fig:pedals}. These sensors detect the distance to the foot using ultrasonic sound waves. The signals were then sent to an Arduino Uno, which had a serial communication with the data collection workstation. The housing cases were modeled in Solidworks and 3D printed.

\subsubsection{Wearable Sensors}
The wearable sensors utilized in this work include a Tobii eye tracking glasses \cite{Tobii2022Tobii3}, a Shimmer physiological sensor \cite{Shimmer2022Shimmer3Unit} and a tactile pressure sensing glove \cite{PPS2022TactileGlove}. The eye tracking glasses is a Tobii Pro Glasses 3, which integrated the micro eye cameras in the lenses to give the wearer an unobstructed view. It also has a scene camera and snap-on lenses for vision correction. The physiological sensor Shimmer3 GSR+ unit provides measurements on galvanic skin response (GSR) and electrocardiogram (ECG). It has a wrist strap connected to two probe electrodes on fingers, and was worn on the non-dominant hand by the participants. The TactileGlove is embedded with 65 individual sensing elements throughout the palm and fingers. They are wireless and battery powered and were worn on the dominant hand of our participants. We followed the CDC guidelines on COVID-19 to disinfect all wearable sensors before each use, and a plastic disposable glove was worn inside the TactileGlove.

\subsection{Experiment Design}
\begin{figure*}
    \centering
    \includegraphics[width=0.7\linewidth]{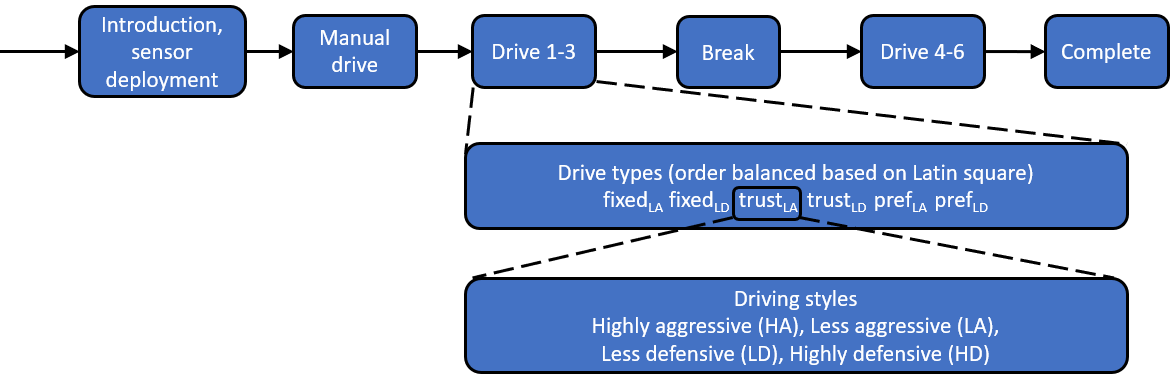}
    \caption{Experiment Protocol}
    \label{fig:protocol}
\end{figure*}

\begin{figure}
\centering
\subfigure[Pedestrian-related event \label{fig:ped}]{\includegraphics[width=.48\linewidth]{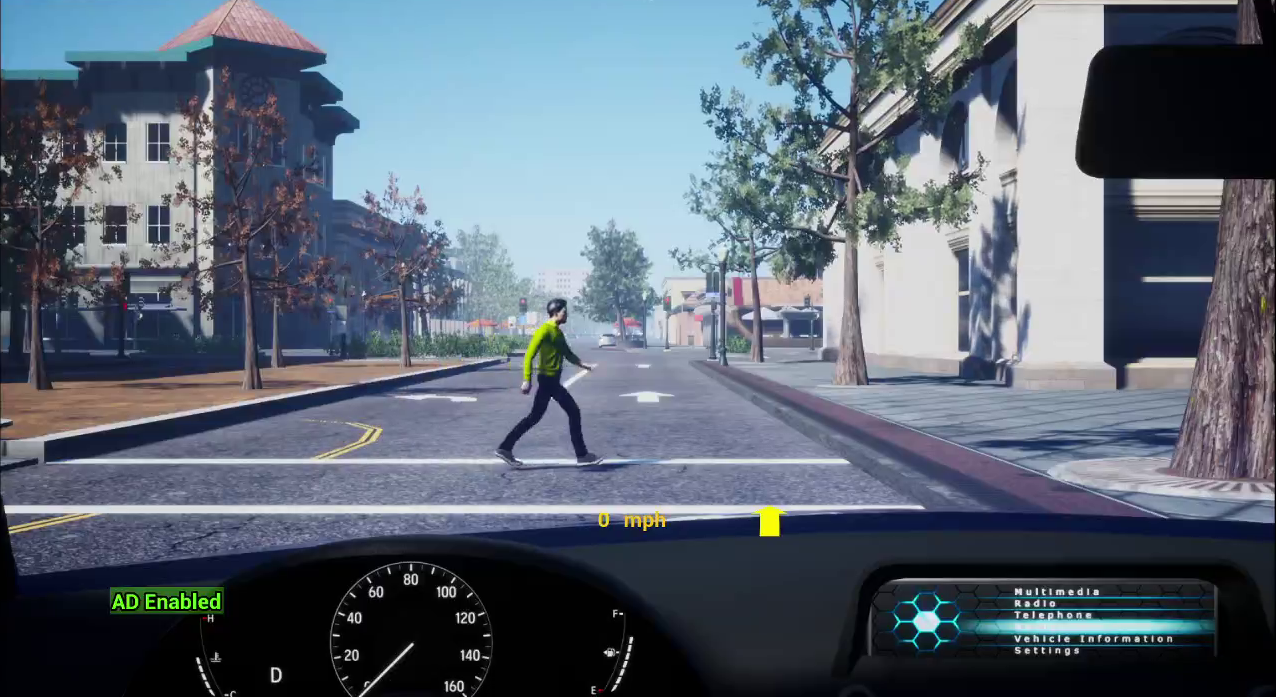}} \subfigure[Traffic-related event \label{fig:car}]{\includegraphics[width=.48\linewidth]{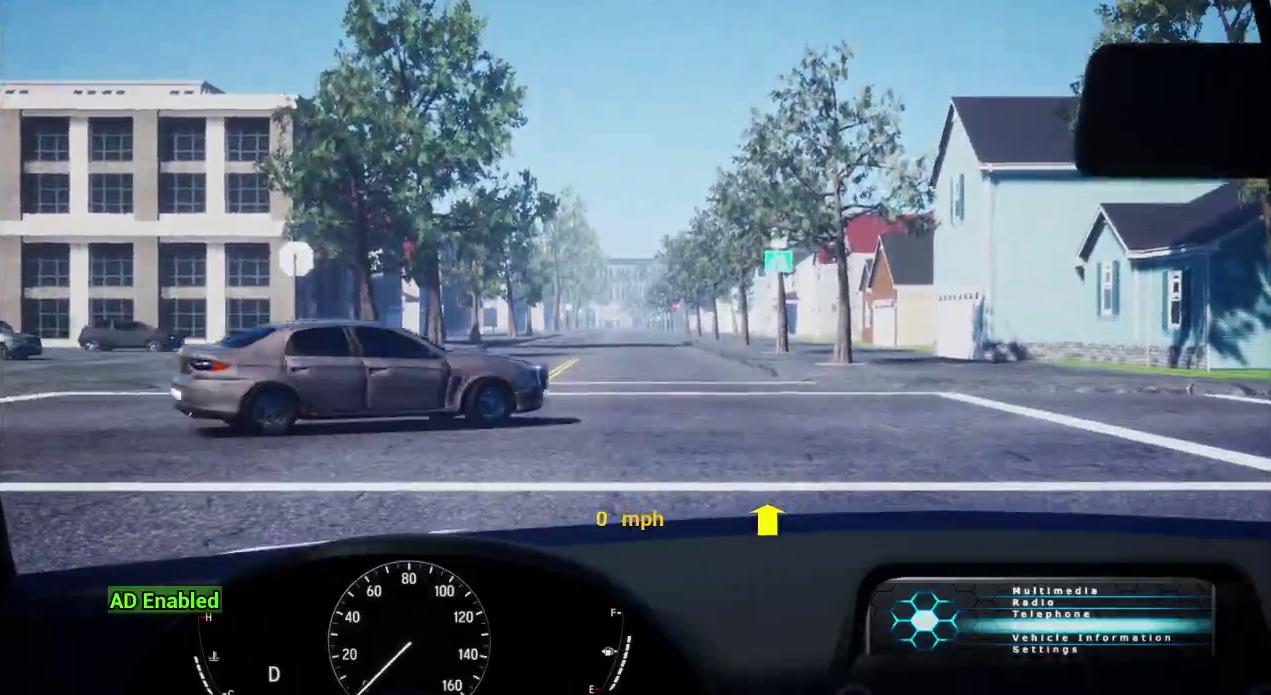}}
\caption{Screenshots of pedestrian and traffic related events}
\label{fig:events}
\end{figure}

The experiments consisted of six driving sessions of six types of driving style adaptations: two fixed and four adaptive. The experiment protocol is shown in Fig. ~\ref{fig:protocol}. In each session, an AV passes through sixteen urban intersections and every other intersection had either of two different event types: pedestrian-related or traffic related, as shown in Fig. ~\ref{fig:events}. Thus, there were a total of eight event intersections per session. Pedestrian related events include pedestrians on the sidewalk, crossing at the crosswalk, at the intersection, and walking at the intersection. Traffic-related events include right turns at a red light, following a lead vehicle, yield and left turns, and a two-way stop. All eight events randomly occurred in each driving session. During the course of automated drives, participants were asked to takeover by pressing the throttle and brake pedals whenever they were feeling unsafe or uncomfortable. Once the participant removed their pedal inputs for a continuous two seconds, the AV would resume control of the car. For a given event, if the participant pressed the pedals, we would annotate the event as takeover presence on brake and throttle pedal, respectively.

We designed four levels of driving styles: 1) highly aggressive (HA), 2) less aggressive (LA), 3) less defensive (LD), and 4) highly defensive (HD). These driving styles vary on driving parameters such as headway, acceleration and minimum distance to decelerate (MDD), as shown in table ~\ref{tab_IDMparams}. The automated driving was implemented using a modified intelligent driver model (IDM) and Stanley controller based on the parameters defined in \cite{Natarajan2022TowardPreferences}.  The two fixed driving style adaptations used the LA and the LD driving styles throughout all events in a driving session. For the four sessions of adaptive driving styles, two included trust-based adaptation and the other two included preference-based adaptation, where the system adaptively chose driving aggressiveness from above-mentioned four levels based on the participant's response.

\begin{table}
\centering
\caption{Parameters used for the IDM controller for the four driving styles.}\label{tab_IDMparams}
\begin{tabular}{lcccc} 
\hline
{Parameter name} & LA       & HA       & LD       & HD        \\ 
\hline
Speed (m/s)             & 13    & 14   & 12    & 11    \\
Max acceleration(m/s2)  & 4     & 5    & 3     & 1   \\
Max deceleration(m/s2)  & 5     & 6    & 2     & 1.5   \\
MDD at intersection (m) & 15    & 20   & 8     & 5 \\
MDD from a pedestrian (m) & 22  & 28   & 15    & 12.5 \\
MDD from a car (m)      & 9     & 8    & 11    & 12 \\
Stop sign duration (s)  & 2     & 1.8  & 2     & 3  \\
\hline
\end{tabular}
\end{table}

In the trust-based adaptive mode, the driving style changed based on a single-item survey that measures the change of trust in the system. We designed these two modes because existing literature found strong correlation between trust in AV and the preferred driving style \cite{Natarajan2022TowardPreferences}. The survey question was presented on the screen and the survey questions present after each of the event. There were five options: greatly increase (+2); slightly increase (+1), stay the same (0), slightly decrease (-1), and greatly decrease (-2). For each selection the participants made, the system recorded the numeric value. Once the accumulative values changed by 2, the driving style would change, where accumulative +2 and -2 would lead the driving style to be more aggressive and more defensive, respectively. We chose the trust change to be a numerical value instead of a continuous one, because we wanted to investigate a classification problem first to prove the concept, instead of a regression problem. Similarly, in the preference-based adaptive mode, the driving style would change based on the preference measurement survey on the screen, and each change from the previous selection would result in a change in the driving style. Specifically, three options were provided on participants' preferred AV driving style: drive more aggressively, stay the same, or drive more defensively. For example, if a participant chose "drive more defensively" when the vehicle driving style was LD, then the AV driving style would drop one level to be HD. Finally, the fixed mode was included as the baseline which presented either the LD or LA driving style. The trust-based and preference-based adaptive drives either started with LD or LA as their initial driving styles. This resulted in six automated driving styles: $\text{fixed}_\text{LD}$, $\text{fixed}_\text{LA}$, $\text{trust}_\text{LD}$, $\text{trust}_\text{LA}$, $\text{pref}_\text{LD}$, and $\text{pref}_\text{LA}$.  

\begin{figure}
\centering
\subfigure[Trust survey \label{fig:trustsurvey}]{\includegraphics[width=.48\linewidth]{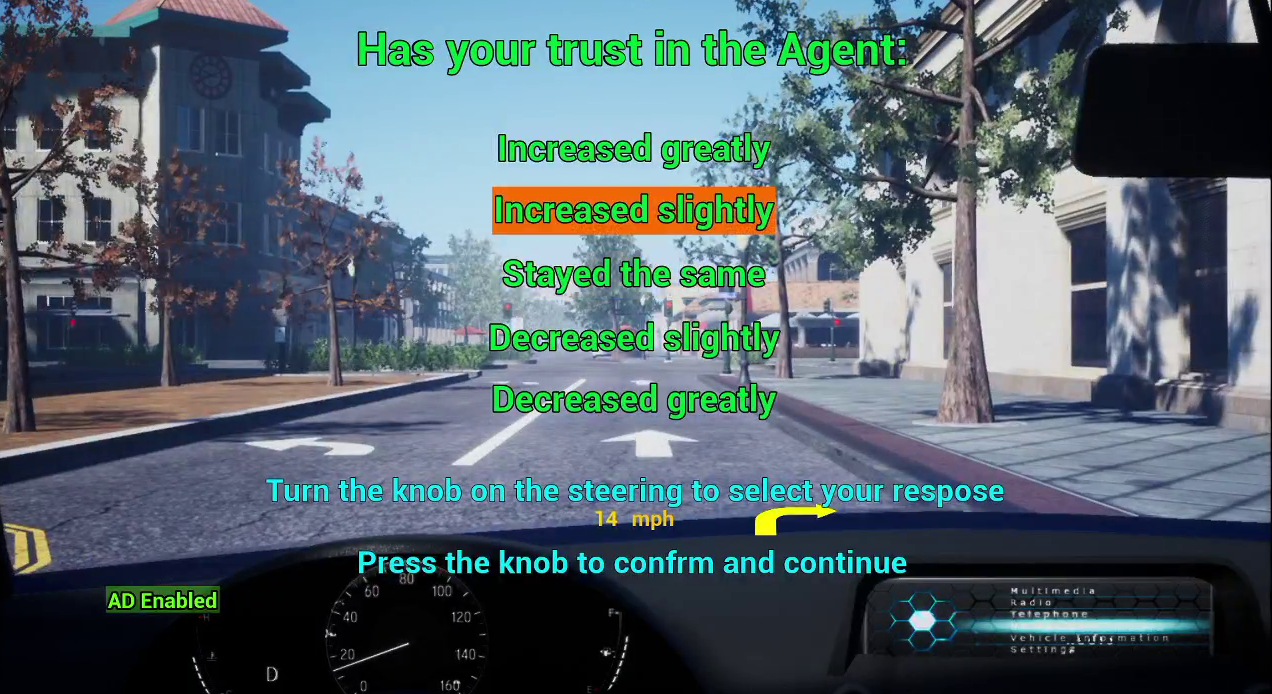}} \subfigure[Preference survey \label{fig:prefsurvey}]{\includegraphics[width=.48\linewidth]{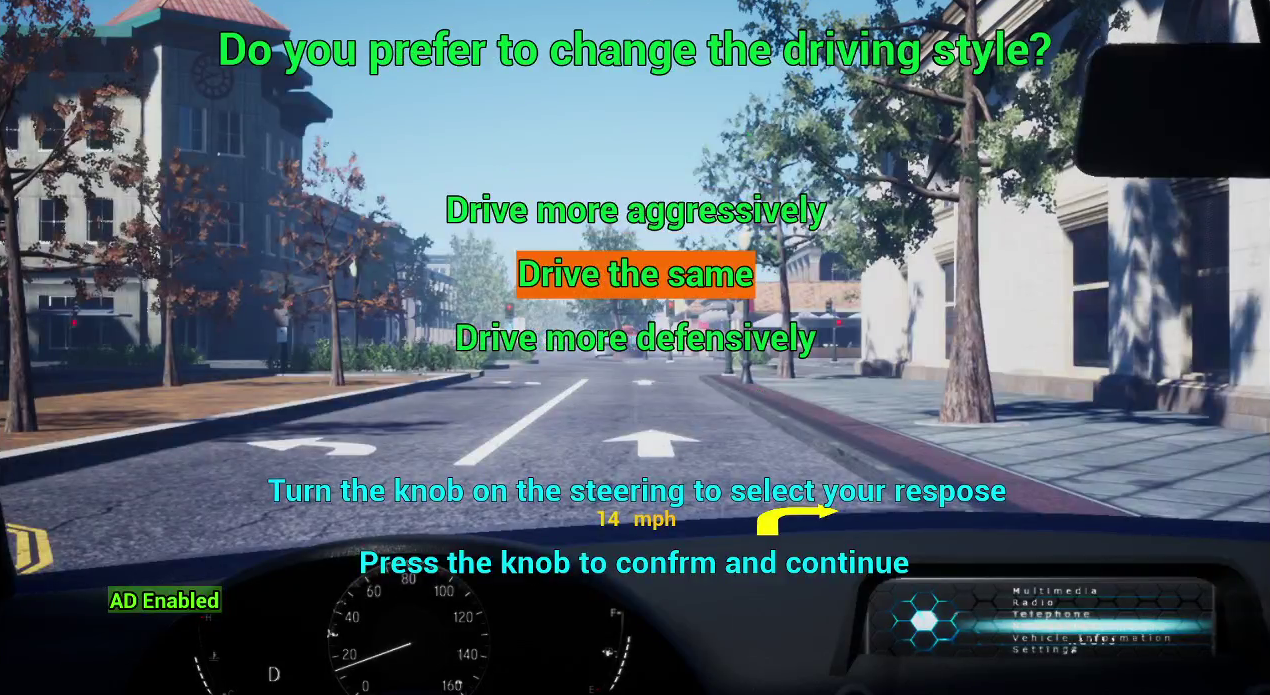}}
\caption{Screenshots of trust and preference questionnaires}
\label{fig:surveys}
\end{figure}

For the experiment protocol, participants were required to sign two consent forms first, including the COVID-19 protection protocols, followed by a demographic questionnaire, which included information such as age, driving behavior, driving history, and their historical use of AV. Participants were then presented with an overview of the experimental procedures, as well as the definitions for trust in automation \cite{Lee2004TrustReliance} and trust in AVs \cite{SAE2018TaxonomyInternational}. 
For the actual experiment, there were a total of seven 10-minute drive sessions, which entailed one manual drive and six automated drive (i.e., SAE Level 2 automation) sets. A 10-minute break was allotted in between the six sets, resulting in the overall experiment duration being approximately two hours. Participants were asked to keep their dominant hand on the steering wheel and their foot on or near the pedals during the experiment. Each automated drive set included two separate driving routes to avoid participants memorizing the driving environment and road events. To avoid drive order adversely influence driver behavioral responses, automated drives were presented in a Latin square counterbalanced order, including two sets in fixed mode (i.e., $\text{fixed}_\text{LA}$ and $\text{fixed}_\text{LD}$) along with two sets for each adaptive mode ($\text{trust}_\text{LD}$, $\text{trust}_\text{LA}$, $\text{pref}_\text{LD}$, and $\text{pref}_\text{LA}$). The manual drive and each automated drive set consisted of all eight events (four pedestrian- and four traffic-related events) in a randomized order. During the automated drive, the simulator would occasionally pause the drive to present questions measuring participants’ real-time trust levels and preferred driving styles. Fig. ~ref{fig:surveys} present screenshots of trust and preference questionnaires, with the response options also listed in the text.  Once all drives were finalized participants were debriefed and the experiment was completed.

\section{Data processing and feature extraction} \label{sec:data_processing}

We ran the experiments on 36 participants, and we used data from 28 participants, whose data did not had any missing data samples. The missing data samples are due to the manual operation mistakes and equipment errors, including eye tracker communication cutting off and distance detection sensor serial communication issue.

We extracted the timestamps of each event and performed pre-processing including cleaning, interpolation and, normalization. For recorded eye gaze, we removed the gaze points where the eye gaze was outside the screen of the simulator. We interpolated the missing data samples in eye gaze and pupil size using nearest-neighbor interpolation; the combined missing data percentage was 6.83\%. Most missing data samples were due to the blinks, and thus most interval of missing data is less than 0.4 seconds \cite{George2017EyeDetection}. We used linear interpolation to replace the missing pedal distance signals. Finally, physiological features and grip forces can vary significantly across individuals \cite{Appelhans2006HeartResponding:}. To eliminate these individual differences, we performed a Z-normalization as shown in \eqref{eqn:normalization}.

\begin{equation} \label{eqn:normalization}
  Z(x)=\frac{x-\mu}{\sigma}\\
\end{equation}

We have extracted a total of 89 features, as shown in Table~\ref{tab:feature}. An asterisk means that we extracted the mean, standard deviation, minimum and maximum values of that feature, within a single event. There are three main categories of data: behavioral, physiological, and situational. The behavioral data include eye gaze, grip force, driving maneuvers and pedal distances, which are the detected driver's physical behaviors. 
The Tobii glasses eyetracker provides real-time gaze location of the participant in the frame of reference of the scene camera mounted on eyetracker. However, we need to identify participants' gaze position in the frame of reference of the driving scene. To achieve this, we used a FLANN Based Matching \cite{Muja2009FASTCONFIGURATION} with SIFT Descriptors \cite{Lowe2004DistinctiveKeypoints} to identify the homography from the scene camera to the three screens of the simulator for each video frame. We then transformed the gaze from the eyetracker scene camera to the driving scene on the three screens. We computed the x and y coordinates of participants' eye gaze, and also the eye gaze velocity as the square root of velocities on the x and y axis. Then, we used a 10 $degrees/s$ threshold \cite{Trabulsi2021OptimizingScreens} to extract eye gaze fixations, dwell time and saccades. We segmented the simulated scenes on the x and y axis into 3x3 regions, and computed the percentage of time participants looked at each region. Then we computed the eye gaze region entropy with \eqref{eqn:entropy}, where \(p_i\) is the percentage of fixation time on the \(i\)th semantic object. For grip force features, we computed the mean, standard deviation, minimum and maximum of it. 

\begin{equation} \label{eqn:entropy}
  H=\sum_{i=1}^{n} p_i\log_2p_i
\end{equation}

We extracted maneuver features from the CAN-Bus signals, including throttle, brake, and steering wheel angles. For the pedal distance detection, we also extracted the distances and approaching behaviors, which are defined as participants moving their feet in the range of 2 centimeters from the pedals. We extracted basic features including current aggressiveness level, event type, and takeovers. 

\begin{table*}
  \caption{Feature List}
  \label{tab:feature}
  \begin{tabular}{cccc}
    \toprule
    Category&Data Modality&Count&Features\\
    \midrule
    Behavioral&Eye gaze&25&$x_{gaze}^*$ $y_{gaze}^*$ $fixation^*$ $dwell^*$ $saccade$, $v_{saccade}^*$ $gaze_v^*$ $entropy_{region}$\\
    &Grip&4&$grip^*$\\
    &Maneuver&12&$throttle^*$,$steering^*$,$brake^*$\\
    &Pedal distance&10&$throttle^*$ $throttle_{valleys}$ $brake^*$,$brake_{valleys}$\\
    \midrule
    Physiological&Pupil&8&$pupil_{left}^*$ $pupil_{right}^*$\\
    &Peripheral&12&$HR^*$ $HRV$ $GSR^*$ $SCR_{count}$ $SCR_{average amplitude}$ $SCR_{max amplitude}$\\
    \midrule
    Situational&Gaze Semantics&15&$p_{objects}$ $entropy_{object}$\\
    &Drive&3&$Aggressiveness$ $type_{event}$ $takeover$\\
  \bottomrule
\end{tabular}
\end{table*}

Physiological data include pupil sizes and peripheral physiological features, which are insidious biological signals that are closely related to human affective states. From peripheral physiological signals, we computed the heart rate (HR) levels and heart rate variability (HRV) from the ECG signals. The HRV is the standard deviation of inter heart beat intervals, which measures how steady the heart is beating. The GSR level measures how well the skin conduct electricity, usually indicating how much the body is sweating. We also extracted the skin conductance responses (SCR), which are the sudden rises of the GSR level. We computed the average and maximum amplitude of the SCR. For pupil sizes, we simply computed the mean, standard deviation, minimum and maximum values of them. 

\begin{figure}
\centering
\subfigure[Original RGB frame \label{fig:RGB frame}]{\includegraphics[width=.7\linewidth]{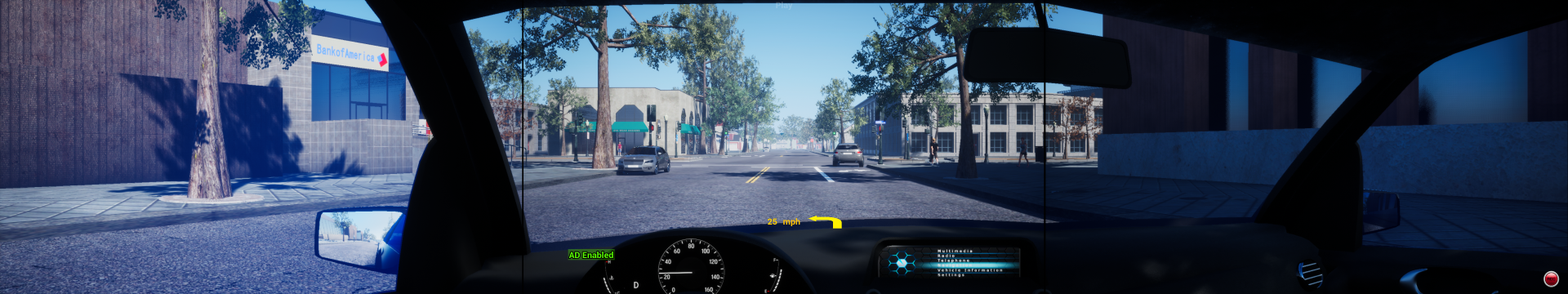}} 
\\
\subfigure[Segmented semantics frame \label{fig:predictedframe}]{\includegraphics[width=.7\linewidth]{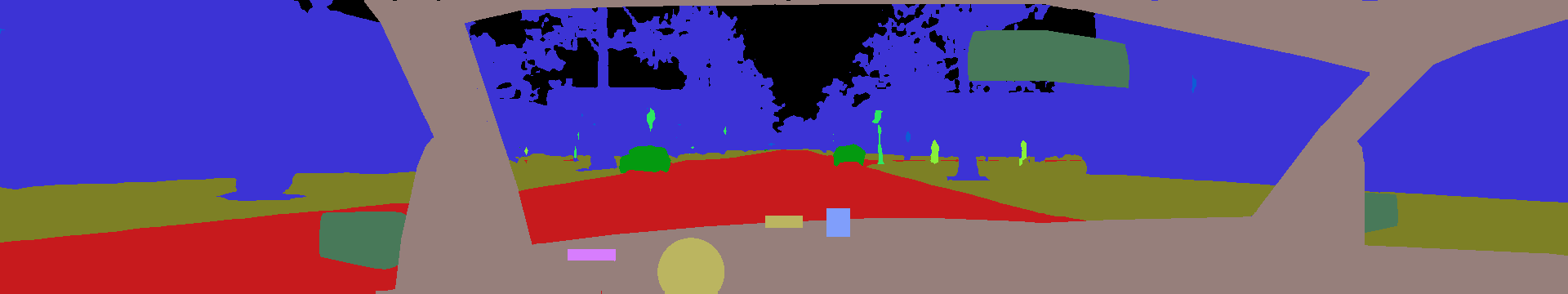}}
\caption{Comparison of scene and segmented frames}
\label{fig:frames}
\end{figure}

The situational category includes gaze semantics and drive features, that are the human perception of the surroundings and the driving task. Eye gaze semantics information, i.e., the type of object the participant is looking at on the scene, is important for driver situation awareness understanding \cite{Hu2021Data-drivenFeaturesb}. We automatically label this semantic labeling for eye gaze data for each frame of the video using semantic segmentation of the driving scene video frames. While in-experiment segmentation of semantics by the simulator is feasible, it requires a lot of computational power and significantly drops the frame rate, which would affect the simulator experience. In order to automatically segment the semantics of each frame after the experiments, we trained a semantic segmentation model using our data. The training dataset included 5,550 RGB frames as samples and their according segmented frames as groundtruth class, which were generated by our driving simulator. We distributed these pictures into training, validation and test sets in a ratio of 8:2:2. We used the DeepLabV3plus library \cite{Chen2018Encoder-DecoderSegmentation} to train the model with a 95.99\% mean intersection over union (MIoU), and applied it to predict semantics information our experimental collected RGB frames. A comparison of the original simulator scene and the predicted semantics segmentation is shown in Fig. ~\ref{fig:frames}. In Fig. ~\ref{fig:predictedframe}, different colors represents different object types on the scene, such as, blue for buildings and trees, brown for inside the car, red for road, and green for other cars. With the eye gaze position on the simulator scenes, we were able to identify what category of object the participants were looking at, and also the percentage of fixation on different objects. We also computed the entropy of fixations of different objects. We performed the feature extraction on each event to form samples, and we excluded the samples with broken data points. The drive features are simply the current aggressiveness of the automated drive, the event type between pedestrian and vehicle-related, as well as the takeover on throttle and brakes. We had a total of 932 complete samples from 28 participants.

\begin{figure*}
\centering
\subfigure[Trust change \label{fig:trustchange}]{\includegraphics[width=0.15\linewidth]{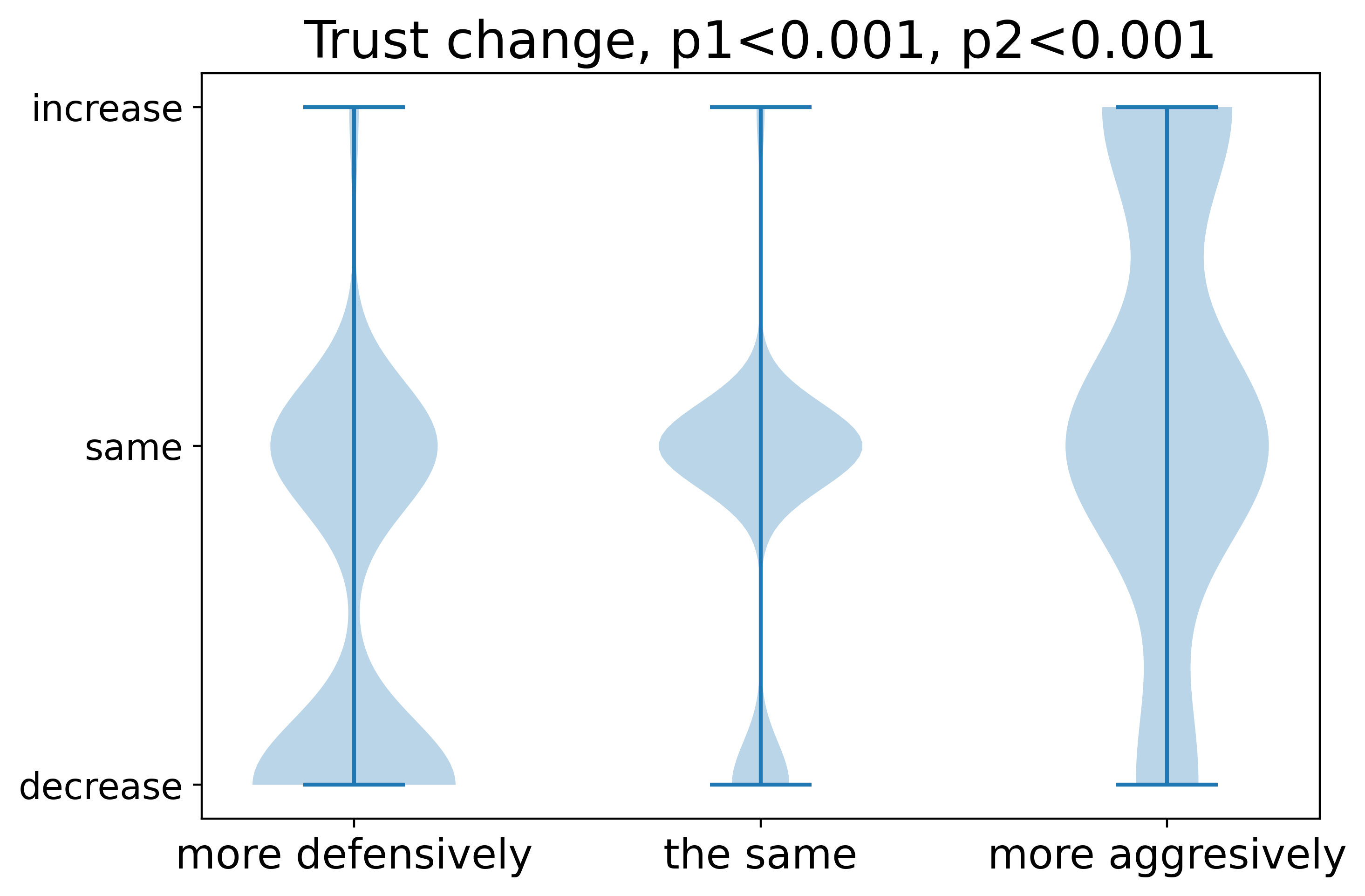}}\subfigure[Gaze y mean \label{fig:gazeymean}]{\includegraphics[width=.15\linewidth]{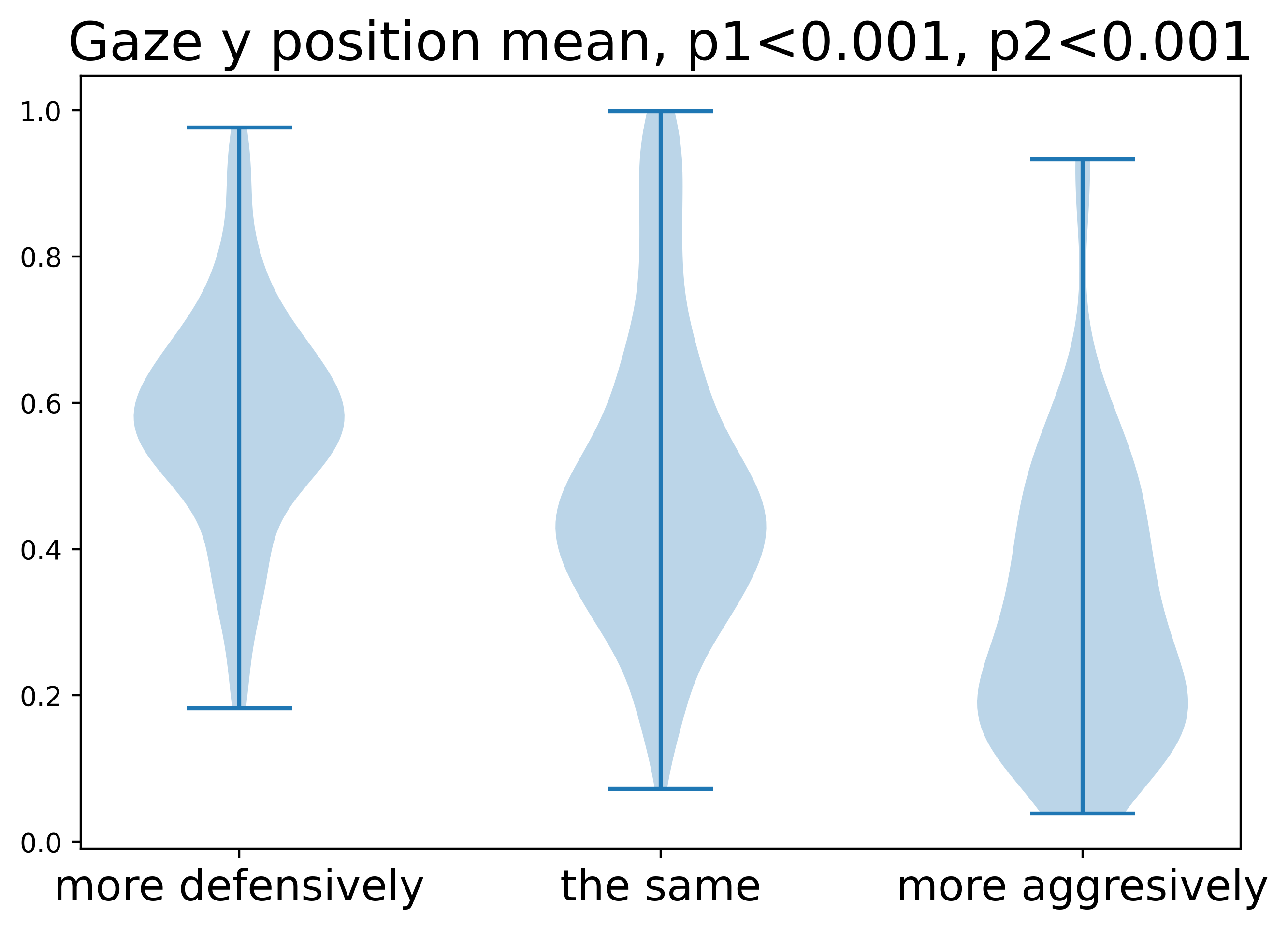}} \subfigure[Pupil std \label{fig:pupilstd}]{\includegraphics[width=.15\linewidth]{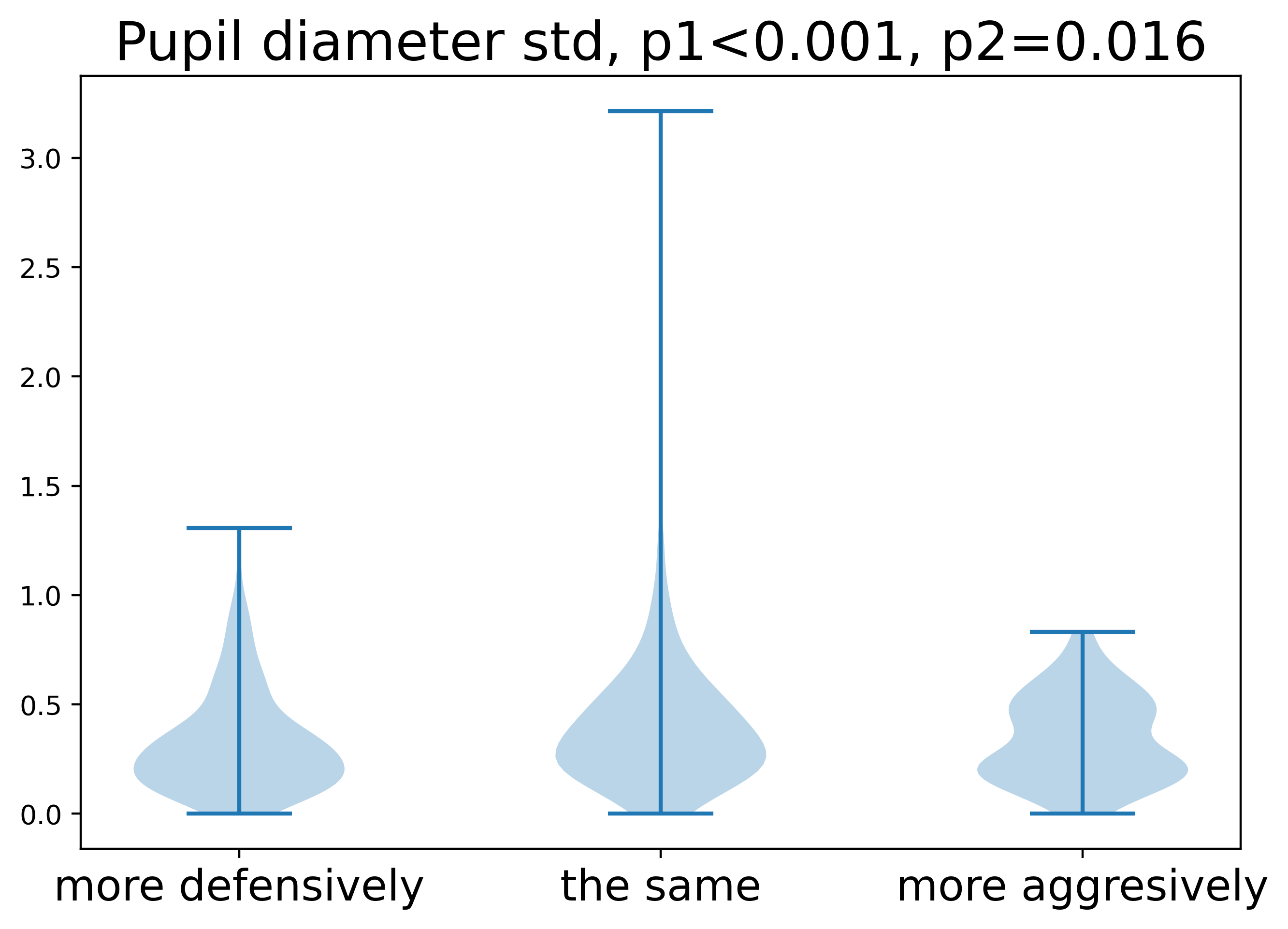}} \subfigure[Gaze percentage on sky \label{fig:sky}]{\includegraphics[width=.15\linewidth]{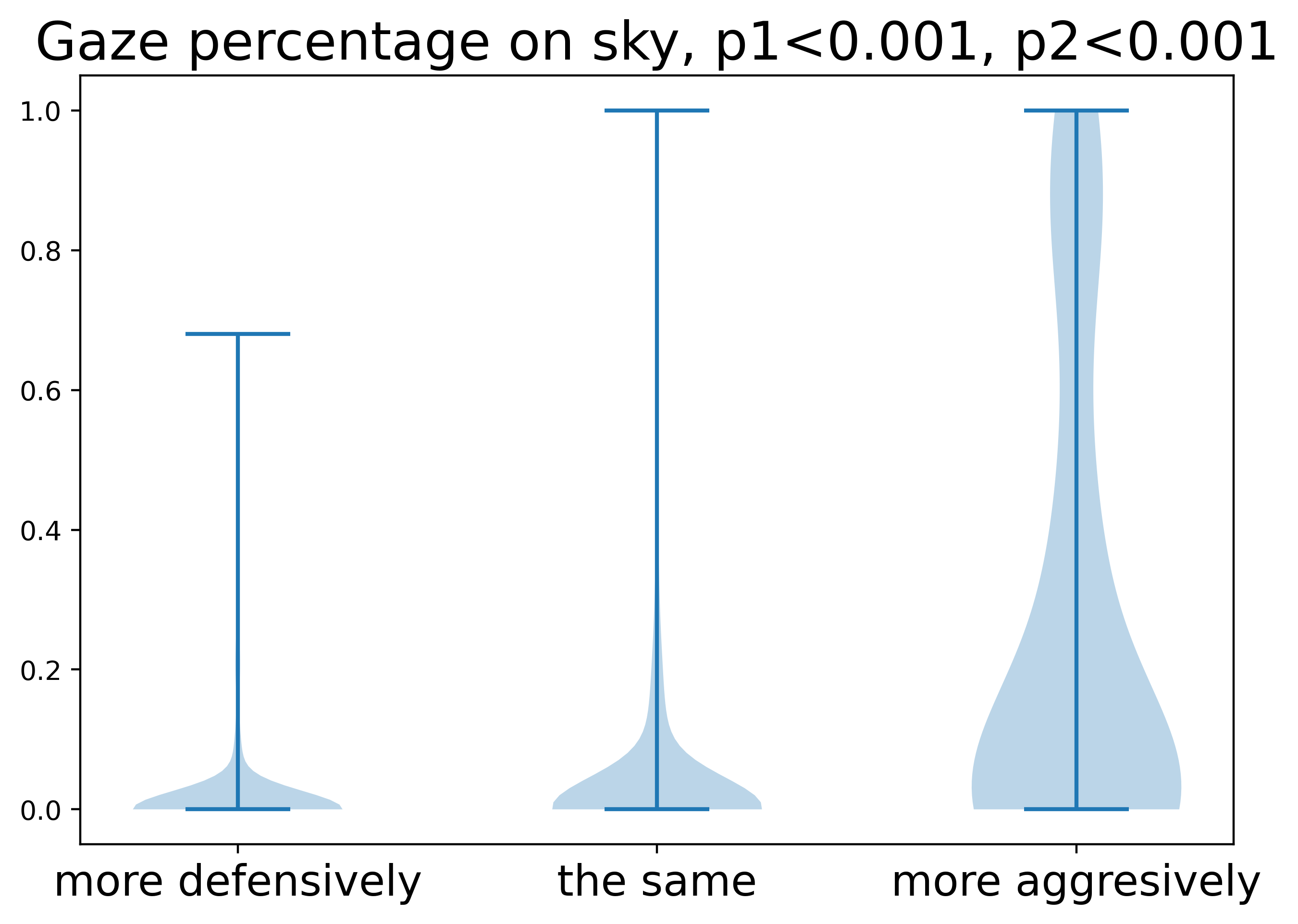}} \subfigure[Gaze percentage on road \label{fig:road}]{\includegraphics[width=.15\linewidth]{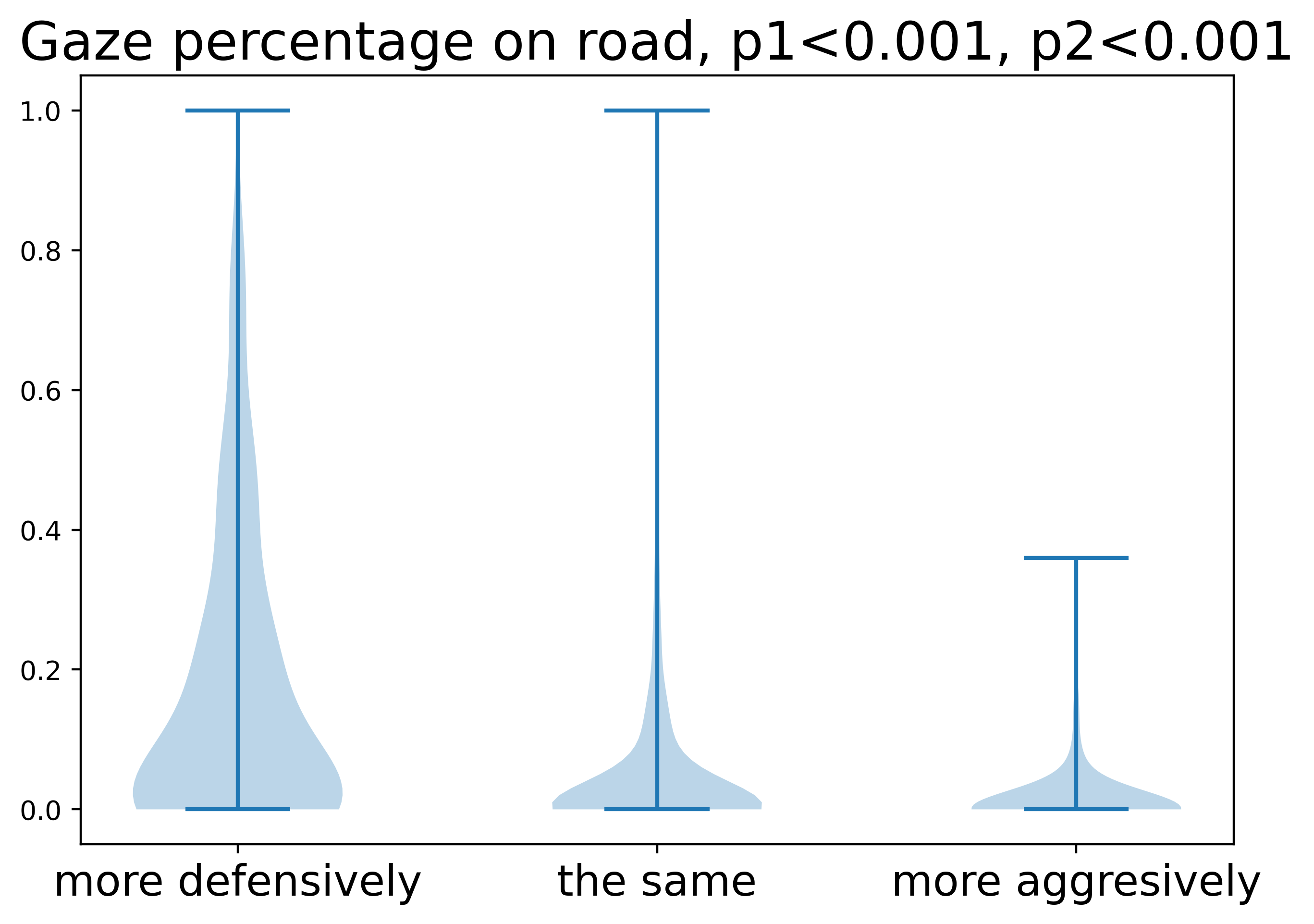}} \subfigure[Gaze percentage on car \label{fig:oncar}]{\includegraphics[width=.15\linewidth]{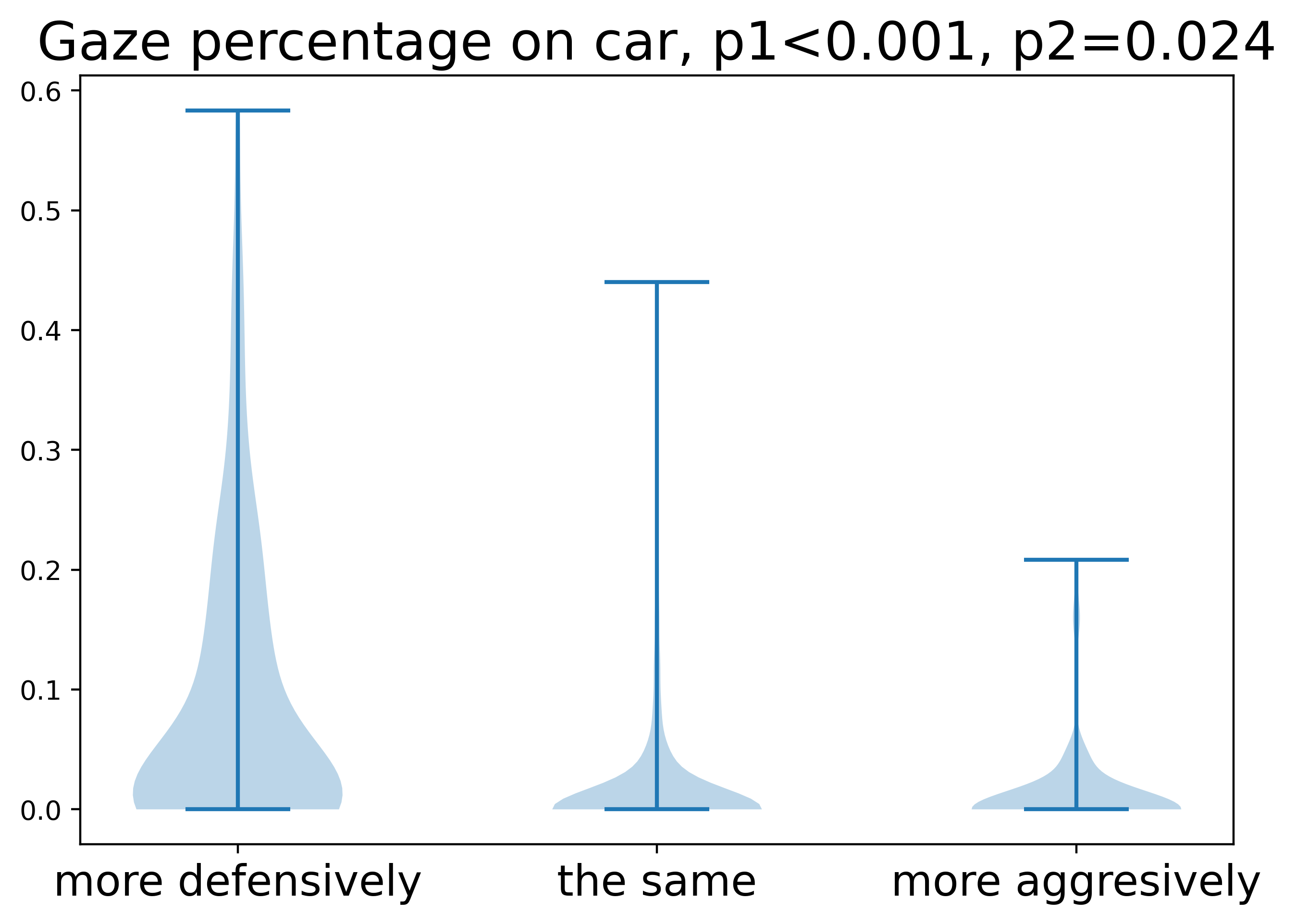}} \subfigure[Object entropy \label{fig:objectentropy}]{\includegraphics[width=.15\linewidth]{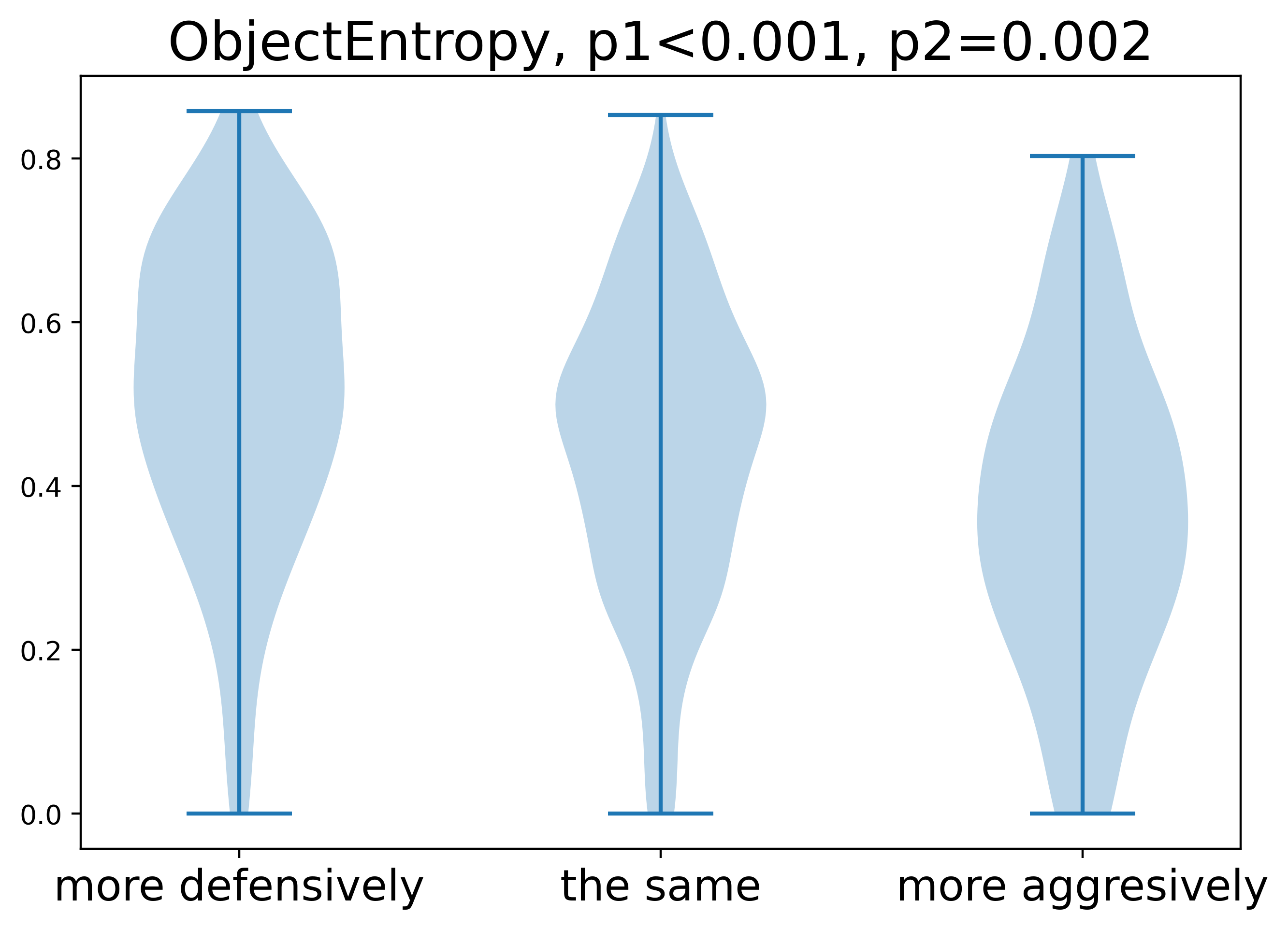}} \subfigure[SCR count \label{fig:scr}]{\includegraphics[width=.15\linewidth]{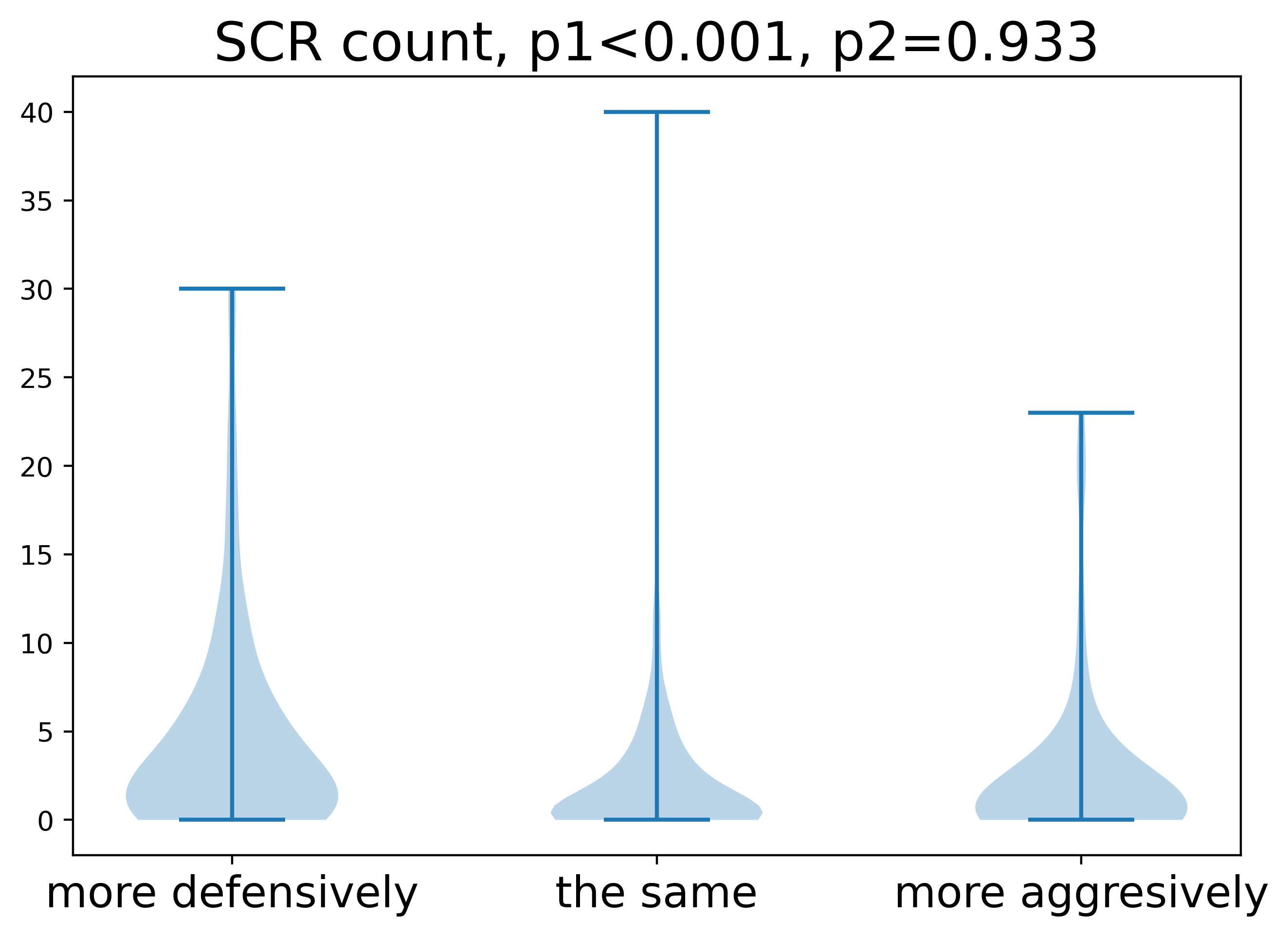}} \subfigure[Brake pedal distance max \label{fig:pedalmax}]{\includegraphics[width=.15\linewidth]{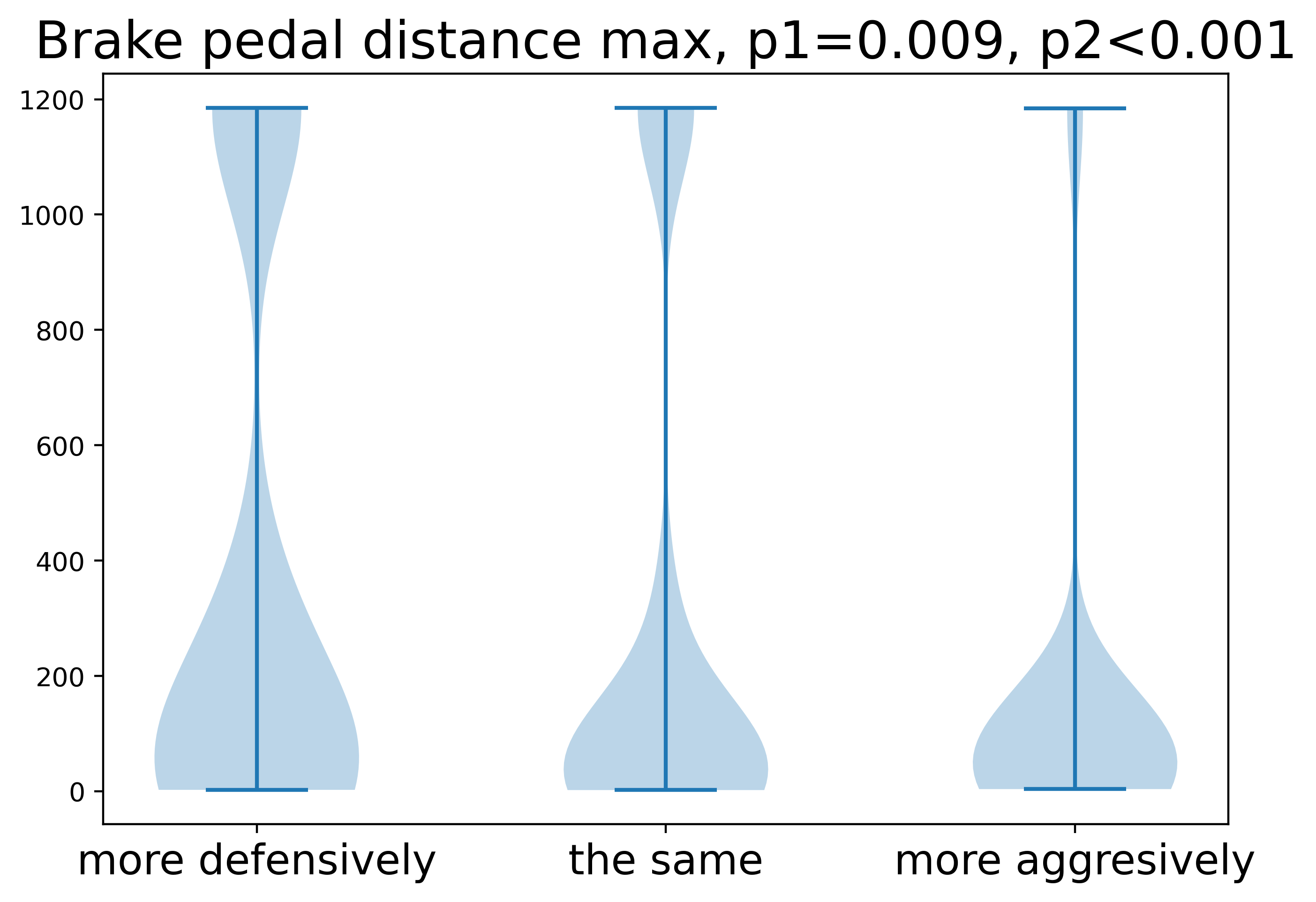}} \subfigure[Brake pedal distance std\label{fig:pedalstd}]{\includegraphics[width=.15\linewidth]{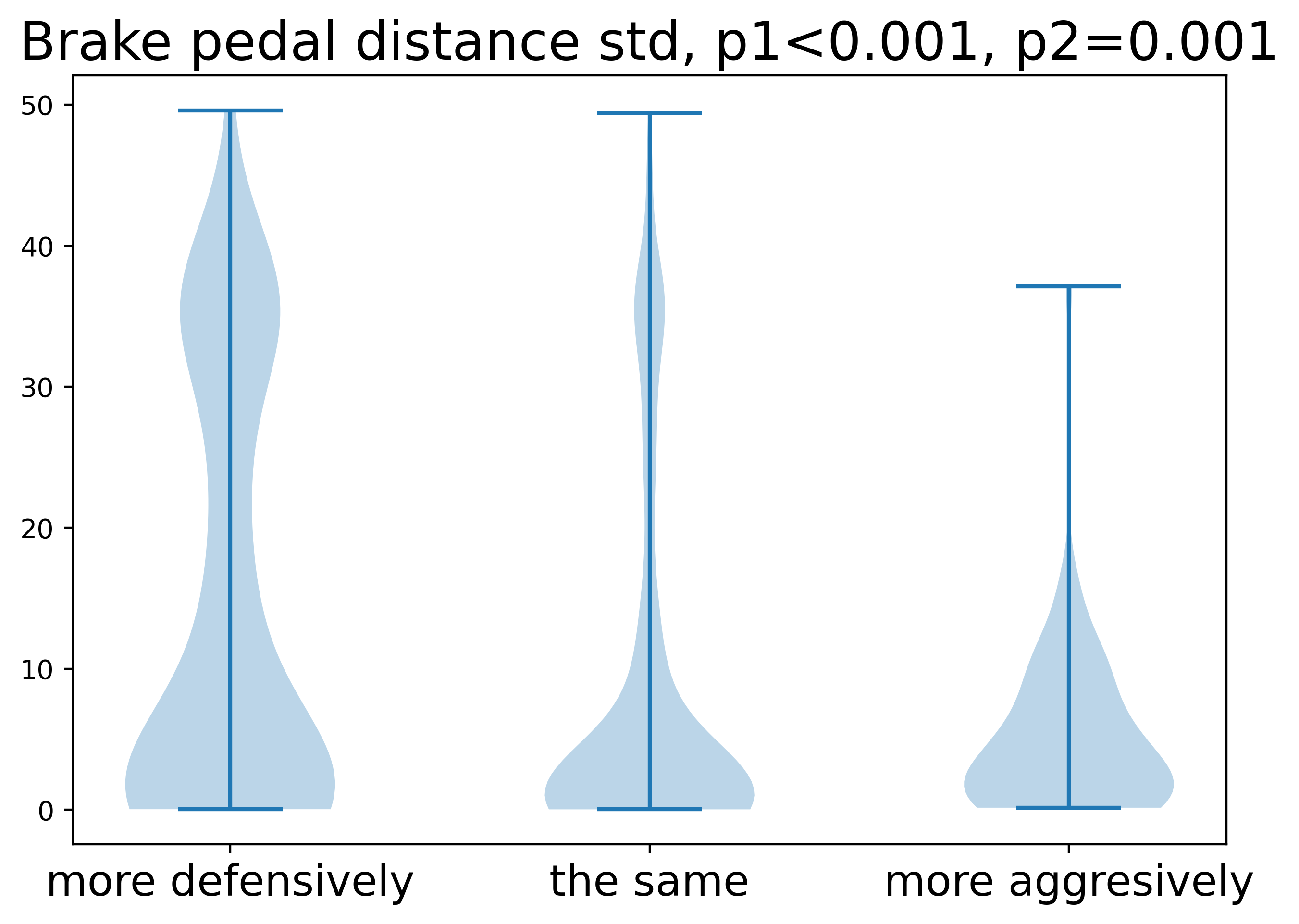}} \subfigure[Grip std \label{fig:gazeystd}]{\includegraphics[width=.15\linewidth]{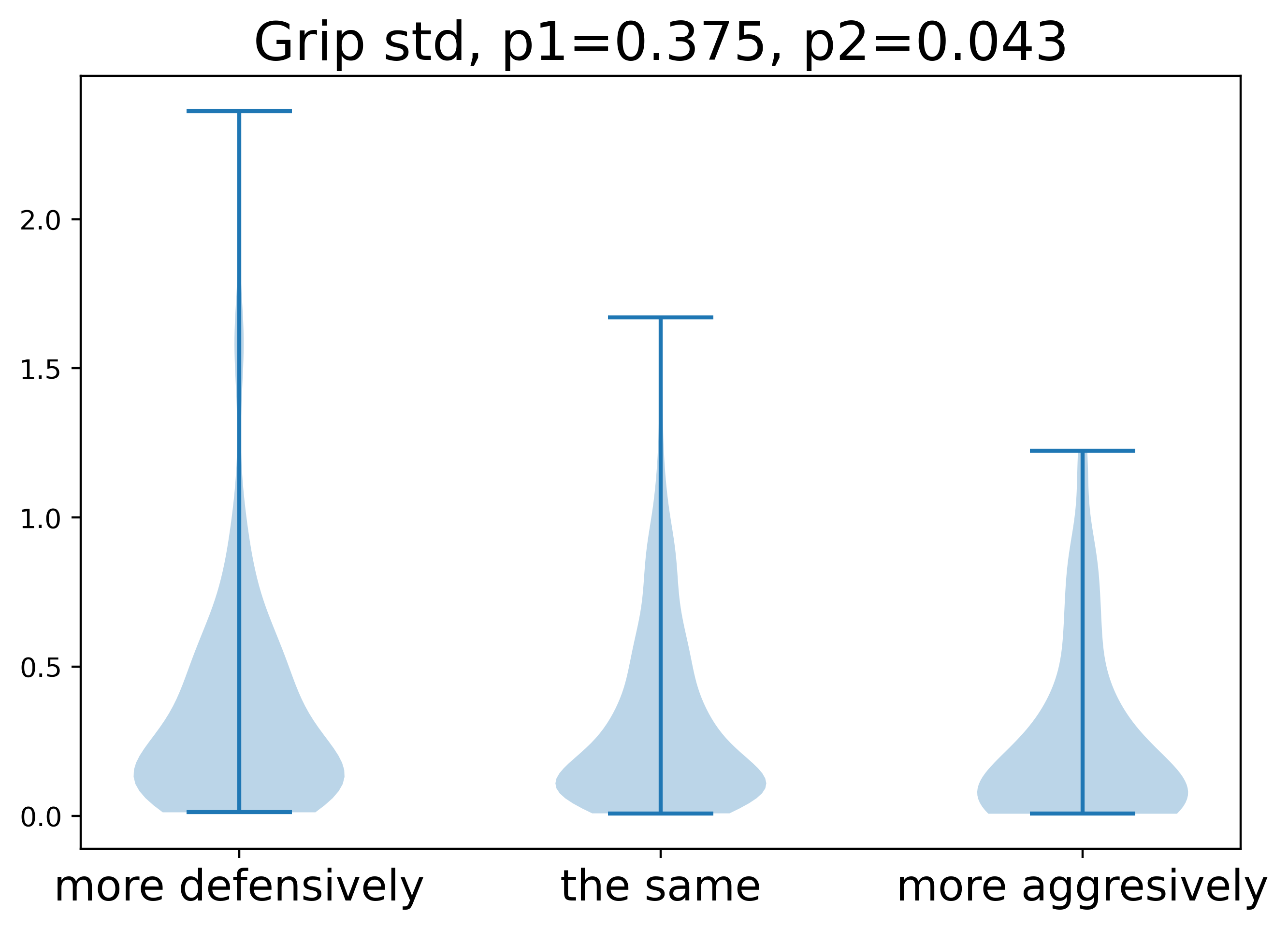}} 
\caption{Statistical Analysis Results}
\label{fig:stats}
\end{figure*}

\section{Statistical analysis and modeling} \label{sec:stat}
\subsection{Statistical Analysis}
We want to understand how driver behaviors varied with different driving style preferences. Our features all follow normal distribution but may have different variances, and different sample sizes. Thus, we ran the Welch's t-test, which is a two sample t-test with the assumption that the samples have unequal variances, and is reliable with varying sample sizes. For each feature, we ran the t-test twice, to compare i) preferring a more aggressive style versus preferring the same driving style; and ii) preferring a more defensive style versus preferring the same style. We present some of the results as shown in Fig.~\ref{fig:stats}, where we found significant differences between preferences. The p-values of the t-tests are included in the figures. First of all, we can see that trust increase is correlated with more aggressive preference, while trust decrease is correlated with more defensive preference. This pattern is consistent with the existing work, that drivers tend to have increased trust in a more conservative system \cite{Forster2018CalibrationStudyb}. For the eye gaze, participants mainly focused on the middle parts of the screen, where the pedestrians, roads and cars were, when they preferred a more defensive driving style. As their preference switched to more aggressive driving styles, they paid more attention to the lower region of the simulator, where the car interior displayed. The pupil diameter had increased standard deviations when the participants preferred a more aggressive style, as disengagement can lead to more frequent dilation \cite{Hopstaken2015TheDynamics}. For the semantics, the participants looked significantly more on the sky when they preferred a more aggressive driving style, and more on the road and car when they preferred a more defensiv e driving style. Participants may need to check the traffic related objects more, such as cars and roads, during more aggressive driving styles and hence prefer a more defensive driving style. In the meantime, they might be zoning out when they focused on the sky, so that they may prefer the vehicle to drive more aggressively for efficiency. One more interesting finding on semantics is that participants had increased object entropy as their preference change toward more defensive driving styles. When participants were scanning through more objects, their cognitive workload may increase to result in a more defensive preference. The SCR counts during more defensive preference were significantly higher, that indicated the presence of stress, arousal or excitement. These emotions are strongly correlated with higher workload on driving \cite{Agrawal2021EvaluatingAutomation}. For the foot distances to the pedals, we found that the maximum distance to the brake pedal is smaller, with a smaller standard deviation when the participants preferred a more aggressive driving style. This combination can be interpreted as participants who constantly laying their foot on the brake pedal were more likely to prefer a more aggressive driving style. For the grip force, the participants had a significant lower standard deviation of grip force, as a steadier grip, when they preferred a more aggressive driving style. We did not observe significant difference between the same and more defensive styles.

\subsection{Machine Learning Modeling}
We want to identify driver preference on driving styles, from multimodal behavioral responses of drivers. Driver preference ground-truth came from survey answers on preference change; specifically, whether they prefer to drive more defensively, prefer to drive the same, or prefer to drive more aggressively. The class distribution is 202:660:70, respectively. For future application, it is very inconvenient for new users to report their driving preference to build the initial model. It is critical to validate our model performance using only some participants' data, to identify other participants' preferences. Thus, we split the samples into 4 folds by participant number and ran cross validations. We used the three folds as training set and the other fold as test set. After train-test splitting, we upsampled the two minority classes in the training set, to reach a balanced class distribution for better performance. We used a random forest classifier, from the scikit\-learn library for this three class supervised classification problem \cite{Pedregosa2011Scikit-learn:Python}.

For machine learning, we first investigated the influence of window length on the model performance. We performed a grid search of feature window lengths on each data modality, ranging from 1, 3, 5, 10 seconds, and full event length, averaging $24.28$ seconds. We built the random forest classifier and picked the best performing window length. The optimal window lengths are 1) 1 sec for gaze, semantics and pupil sizes; 2) full event length for physiological features; 3) 3 seconds for grip force; and 4) 10 seconds for CAN-Bus, pedal, grip and drive features. The optimal window sizes may indicate how much time the participants need to form a decision toward driving style preference.   

We selected the optimal feature lengths, and then trained and fine tuned the model for best performance. We ran the 4-fold cross participant validation and the average accuracy of 76.02\% was achieved for the model reached. 

\begin{figure}
  \centering
  \subfigure[Two step training diagram \label{fig:twostep}]{\includegraphics[width=0.45\linewidth]{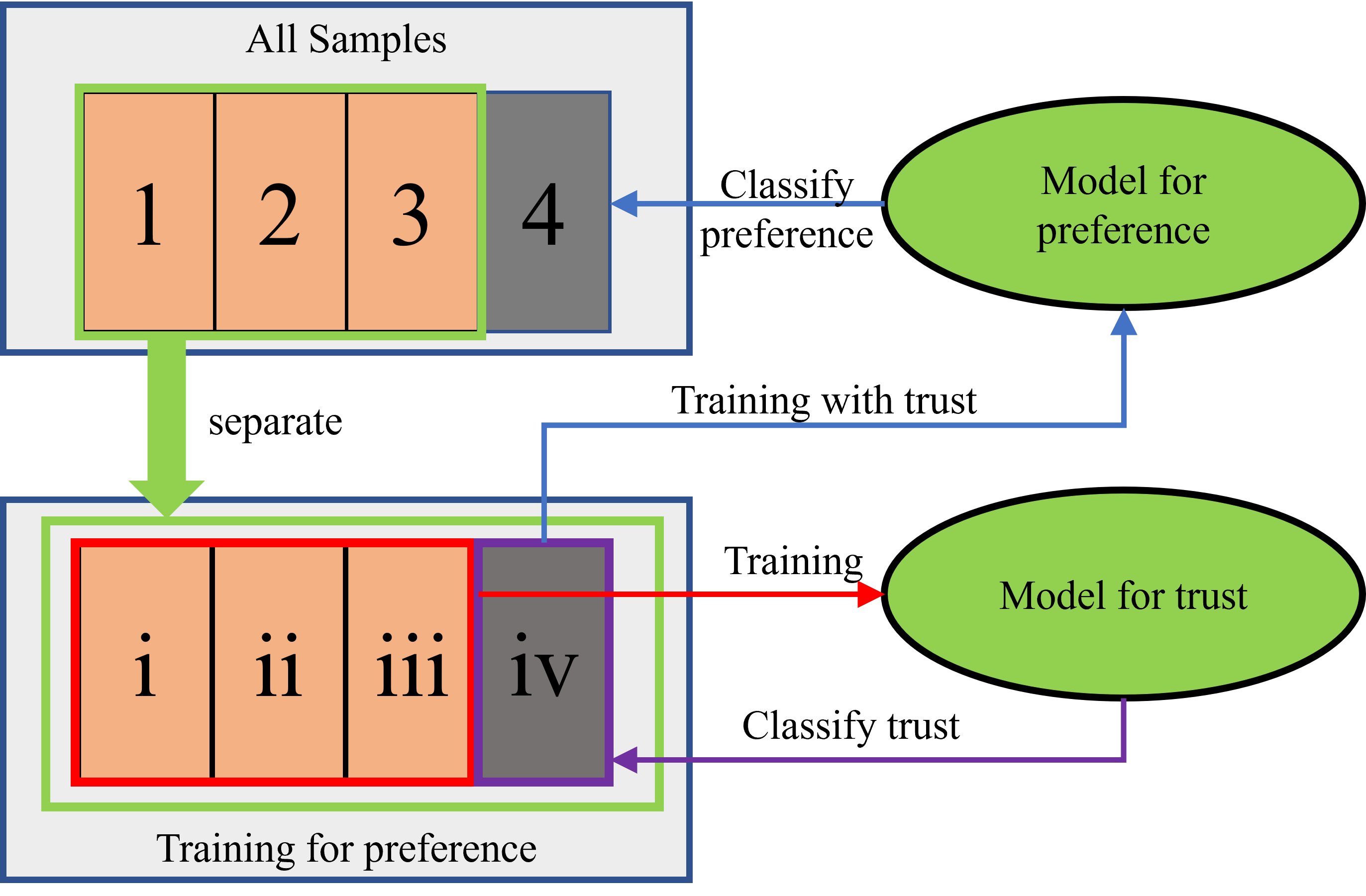}} \subfigure[ROC Curves \label{fig:roccurves}]{\includegraphics[width=0.5\linewidth]{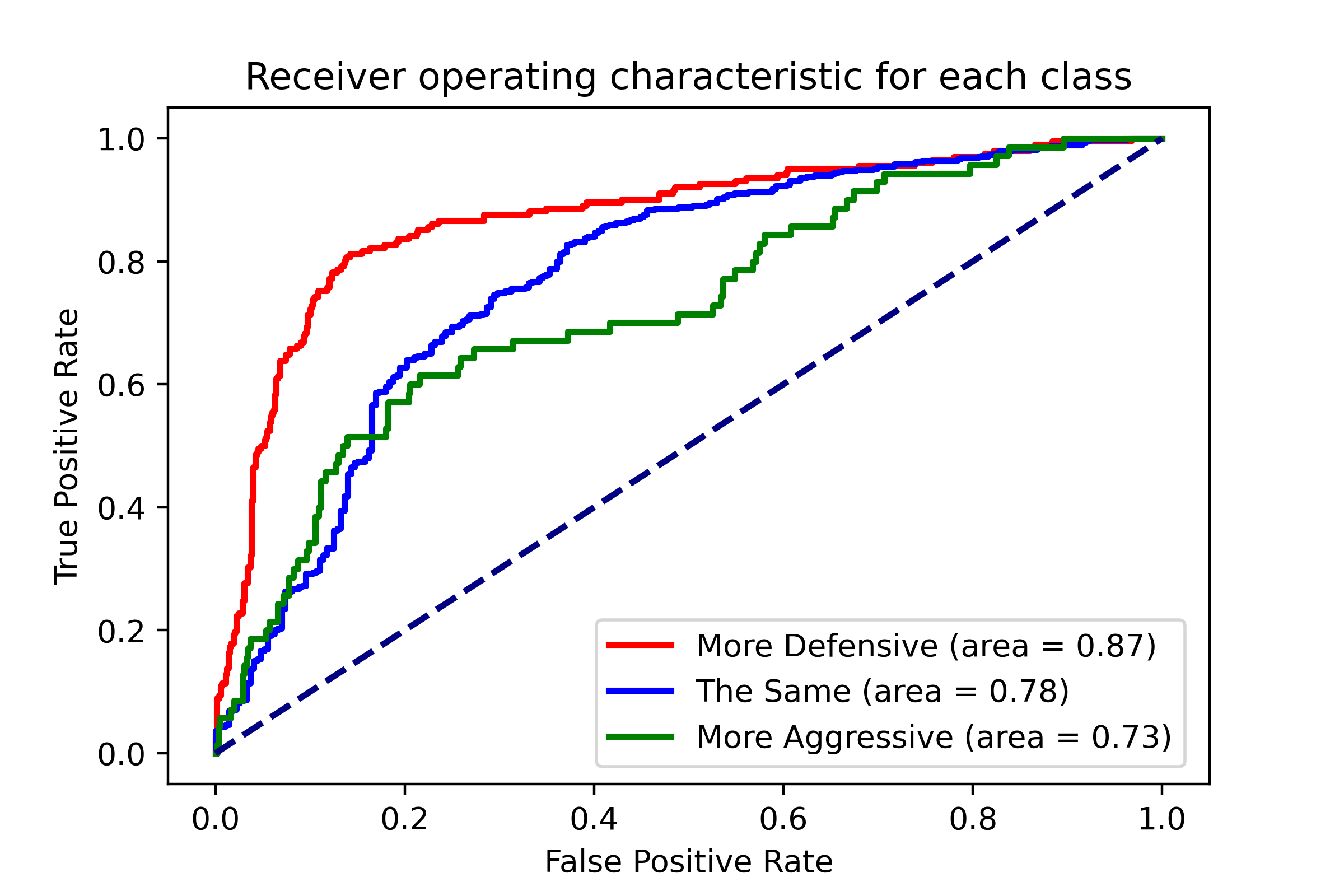}}
  \label{fig:ml}
  \caption{Training pipeline and results}
\end{figure}

Trust is an important factor toward preference in automated drives and machine learning detection of it has been previously studied \cite{Akash2018AGSR}. Thus, we can use the trust as an intermediate factor for preference identification, as shown in Fig.~\ref{fig:twostep}. For the cross validation for preference identification model, we separated the training data into another 4 folds, and we used three participant folds as training data for the trust model. The trust model then classified trust changes and levels on the fourth fold. In this case, we added trust labels from multimodal data, without knowing explicit trust inputs from the test participants. The accuracy for trust change and trust level classifications are 71.22\% and 86.56\%, respectively. Finally, we push all the training data with trust labeling to the preference model. The two step training pipeline increased the accuracy of preference identification to 77.16\%. We also assessed the model with One\-versus\-Rest (OvR) receiver operating characteristic curve (ROC), as shown in Fig. ~\ref{fig:roccurves}. The area under curve (AUC) values for each preference reached 0.87, 0.78 and 0.73, respectively. 

\subsection{Ablation Study}
\begin{figure}
  \centering
  \includegraphics[width=0.8\linewidth]{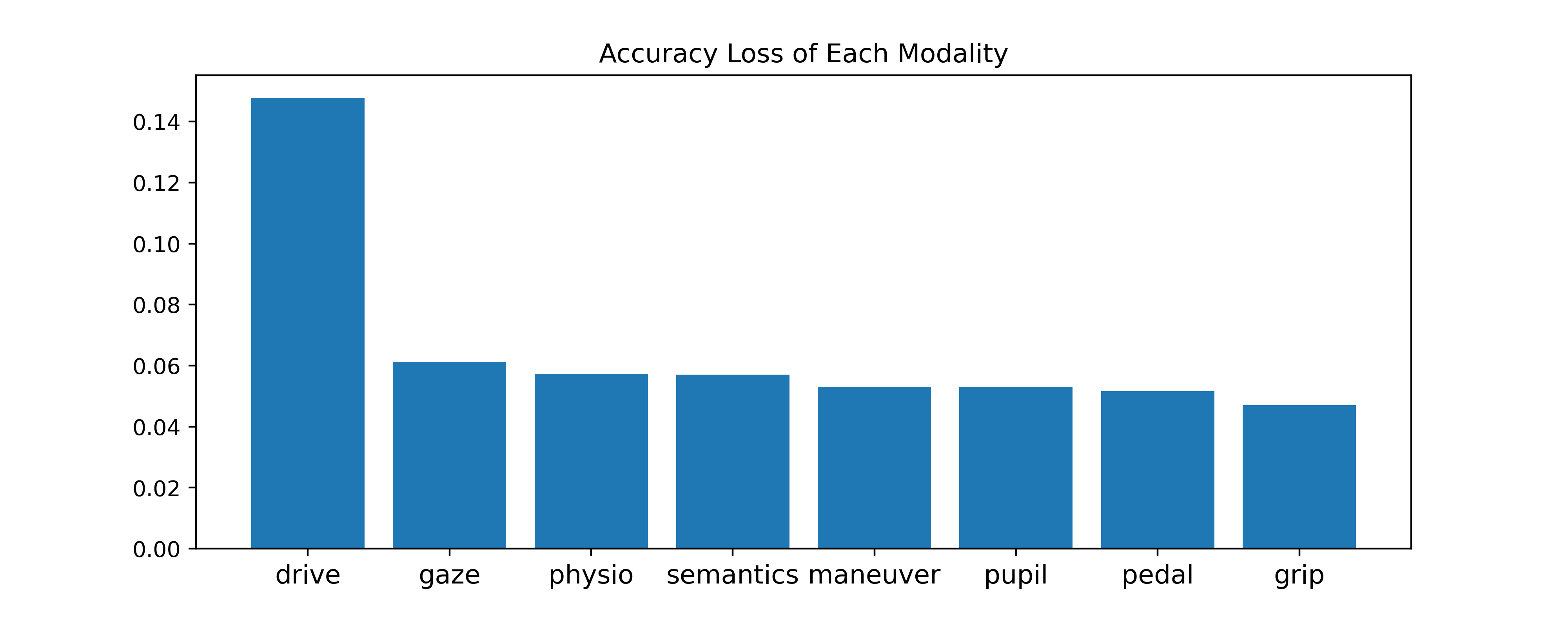}
  \caption{Ablation Study Results}
  \label{fig:ab}
\end{figure}

We wanted to assess which features were the most informative towards driving style preference. We did an ablation study, where we leave each data modality out and find out the model accuracy loss. The results are shown in Fig. ~\ref{fig:ab}. The most dominant modality is the drive information, such as the current aggressiveness level and takeover behaviors. The accuracy losses for multimodal sensing modalities were more evenly distributed, with top important features being gaze, physiological, and semantics data. We also utilized the sequential feature selection, that formed feature subsets of 10 in a greedy fashion. The 10 most informative feature subset include the mean, standard deviation and minimum of the gaze y coordinate, the maximum of the gaze speed, the standard standard deviation of the right pupil, the standard deviation of GSR, the standard deviation and maximum of throttle, the approaching count on the throttle pedal, and the standard deviation of the grip. The most informative features are quite evenly distributed across different modalities once again. These results together may suggest that there may not be a single dominant feature for driving preference identification, and multimodal data may provide a more robust performance.

\section{Conclusion} \label{sec:conclusion}
We presented data collection, analysis and machine learning of driving style preference on SAE L2 automated vehicles. The multimodal data include implicit data modalities, including peripheral physiological signals, eye gaze, CAN-Bus signals, grip force, and pedal distances. These measurements do not need user inputs on driving style preferences, so that they could focus on their driving tasks or other goals during the drives. Based on collected data, we discovered behavioral patterns for driving style preference, such as increased fixation on road, increased object entropy and increased SCR when they preferred a more defensive driving style, as well as higher pupil size std, more gaze fixation on sky, and steadier grips when they preferred a more aggressive driving style. We extracted features with optimized window sizes, then we trained machine learning models to identify driving style preference, with a cross-validation process to only utilize other participants' data as training to have a more generalized result. The promising identification performance indicate that it is feasible to detect driver's preference on driving styles with implicit inputs, with pre-trained models from other drivers. The best model we trained had an accuracy of 79.86\%. We also analyzed the feature importance of our multimodal data through a sequential feature selection approach. The results suggest that the variety of wearable sensing may have augmented the identification. Most of the sensors used in this study is wearable or portable on a vehicle, and the computational power required to make identifications in real-time is not significant. Therefore, this study is a step towards future applications of real-time on-vehicle adaptive driving styles through wearable sensing. Such driving styles with implicit inputs would significantly improve user comfort and acceptance on ADAS features or automated drives.

The major limitation of our work comes from the small amount of data we were able to collect due to the extensive amount of time for data collection. With more time and resources, we could potentially expand our dataset size to improve our model performance and robustness. In addition, we will be investigating deep learning-based algorithms, to get improved performances in detection. Another future direction is to investigate sparse data preference survey timings. Although we conducted survey after intersection events, it is not clear which exact event triggered the preference change. We can further investigate time-series relationships between the multimodal behavioral responses and events, to locate the factors contributing to preference change. Despite these limitations, the work is novel to identify preference on automated vehicle driving style through implicit multimodal data. This work prove the concept of using wearable and on-vehicle sensing to create adaptive driving styles in automated driving, to improve broaden user acceptance and comfort.


\bibliographystyle{ACM-Reference-Format}
\bibliography{Mendeley}

\end{document}